\documentclass[pra,aps,superscriptaddress,secnumarabic,eqsecnum,showpacs]{revtex4}
\usepackage{epsfig,amsmath}

\begin{document}

\title{Quantal Two-Centre Coulomb Problem treated by means of the
Phase-Integral Method III. Quantization Conditions in the General Case
expressed in Terms of Complete Elliptic Integrals. Numerical
Illustration}

\author{N. Athavan} 
\thanks{Present address: Department of Physics, 
Government Arts College, Ariyalur - 621 713, India.}
\affiliation{Centre for Nonlinear Dynamics, Department of Physics, 
Bharathidasan University, Tiruchirapalli 620 024,India}

\author{N. Fr\"oman} 

\affiliation{Department of Theoretical Physics, University of Uppsala,
Box 803, S-751 05 Uppsala, Sweden}

\author{M. Lakshmanan} 

\affiliation{Centre for Nonlinear Dynamics, Department of Physics, 
Bharathidasan University, Tiruchirapalli 620 024,India}

\begin{abstract}
In this paper we take up the quantal two-centre problem where the
Coulomb centres have arbitrary positive charges. In analogy with the symmetric
case, treated in the second paper of this series of papers, we use the 
knowledge on the quasiclassical dynamics to express the
contour integrals in the first- and third-order approximations  of the 
phase-integral quantization conditions, given in the first paper of this series
of papers, in terms of complete elliptic integrals. For 
various values of the distance between these charges the accuracy of the
formulas obtained is illustrated by comparison with available numerically
exact results.

\end{abstract}
\pacs{PACS Numbers: 03.65.Sq, 31.15.-p, 31.15.Gy}
\maketitle
\section{Introduction}
\label{sec1}
In this third paper in a series of papers concerning the two-centre 
Coulomb problem we take up the general case in which the positive charge
numbers $Z_1$ and $Z_2$ of the two Coulomb centres are arbitrary. The study of such general systems 
is of considerable importance in the field of molecular physics and elementary particle physics.
For example, the calculation of eigenvalues and electronic wave functions
for one-electron diatomic molecules with fixed internuclear separation is the starting
point for an accurate description of molecular vibrations and rotations and of
ion-atom scattering \cite{meli,hat,tin,win}. Also the calculation of radiative transition
probabilities for a $\pi^-$ meson
moving in the Coulomb field of two fixed nuclei \cite{pon1} is a good example of
the two-centre Coulomb
problem dealt with in elementary particle physics. Such calculations are of physical interest in
connection with experimental \cite{Dun,char,Dun1} and theoretical\cite{pon2} research on the absorption
of $\pi^-$ mesons stopped in substances containing hydrogen. 

In the second paper in this series \cite{ath2} the symmetric case,
$Z_1=Z_2$, was considered. Using for the two-centre Coulomb
problem the general phase-integral quantization
conditions  derived in the first paper \cite{ath1}, the relevant
contour integrals for the first- and third-order approximations were expressed 
in terms of complete elliptic integrals
so that numerical evaluation of energy eigenvalues and separation constants 
can easily be carried out. 
The evaluation of the various contour integrals was facilitated through
suitable transformations of the $\xi$- and $\eta$-variables, which can be
related to the quasiclassical motion of a particle. 
In the symmetric case, $Z_1=Z_2$, the square of the base
function $Q^2(\eta)$ has a symmetry, as discussed in \cite{ath2}.
As a result of this, the evaluation of the quantities $\alpha =\beta$, $L$, $L'$ and  $\bar K$
pertaining to the $\eta$-quantization conditions were performed with the use of particular transformations
in a rather simple way, while for the quantity $\tilde L$ in the $\xi$-quantization condition
rather general transformations are necessary.
However, in the general case, where $Z_1$ may be different from $Z_2$,
no such symmetric structure exists for $Q^2(\eta)$, and both $Q^2(\eta)$ and
$\tilde Q^2(\xi)$, given by eqs.(3.2a,b) in \cite{ath1}, can have, 
besides the poles, the  following structure of the zeros:

\begin{enumerate}
\item
Case $\Lambda=|m|\ne 0$:
\begin{enumerate}
\item
Four real zeros 
\item
Two real and two complex conjugate zeros 
\end{enumerate}
\item
Case $\Lambda=0$:
\begin{enumerate}
\item
Two real zeros 
\item
Two complex conjugate zeros 
\end{enumerate}
\end{enumerate}
As a consequence, the evaluation of the contour integrals in the cases
of the $\xi$-equation and the $\eta$-equation are analogous, except
that certain coefficients change in the different subcases. This allows
one to use the ``universal'' functions $H^{(2n+1)}$ and $\bar
H^{(2n+1)}$, $n=0$ or $1$, valid for the $(2n+1)$th-order
approximation, which were defined in sec. 3.1.2 and sec. 3.2.2 of
\cite{ath2}. These functions depend on five parameters
$\nu_1,\nu_2,\nu_3,g$ and $k$, which are expressed in terms of the
zeros of either $Q^2(\eta)$ or $\tilde Q^2(\xi)$, and on the parameter
$C$ and $\tilde C$ in the base function $\tilde Q(\xi)$ or $Q(\eta)$,
respectively; see eqs. (3.2a,b) in \cite{ath1}.  Different choices of
these parameters allow one to express the relevant quantities $\alpha,
\beta$ (which may now be different from $\alpha$), $L,L',\bar K$ and
$\tilde L$ in the first- and third-order phase integral approximation
(apart possibly from a sign or a constant factor) as the appropriate
``universal'' function or its real or imaginary part with the
appropriate parameters inserted. When the contour integrals are
evaluated, one can solve the quantization conditions to obtain the
energy levels accurately.

In principle  one can specialize the results of the general case when $Z_1$ may be different from  $
Z_2$ to the particular case when $Z_1=Z_2$ in order to obtain the results of \cite{ath2},
but in practice this is cumbersome due to the different transformation formulas
used for the $\eta$-part of the quantization conditions
in \cite{ath2}. For the $\xi$-part such a specialization implies only that $Z_1+Z_2$
is replaced by $2Z_1$ but no essential simplification.

The plan of the present paper is as follows. In sec. 2 we express
$\alpha,\beta,L,L',\bar K$ and $\tilde L$ in terms of the ``universal''
functions appropriate for the subbarrier case with $\Lambda=|m|\ne 0$.
Then a similar treatment is given for the superbarrier case.  In sec. 3
an analogous procedure is applied to the case $\Lambda=0$.  Finally, in
sec. 4 a detailed numerical analysis of the phase-integral quantization
conditions is carried out for $Z_1 =1$ and different values of $Z_2$
(=2, 5 and 8), and comparision is made with  existing numerically exact
results for the energy and the reduced separation constant.

\section{Case $\Lambda = |\lowercase{m}|\ne 0$}
In this section we utilize for the
case of four zeros of $Q^2(\eta)$ or $\tilde Q^2(\xi)$
the ``universal'' functions $H^{(2n+1)}$, $n=0$ or $1$, introduced in \cite{ath2} and 
related to the $(2n+1)$th-order contribution,  
and  the ``universal'' functions $\bar H^{(2n+1)}$, $n=0$ or 1, which will be
introduced in sec. 3.2.2. In the different cases one finds,
as explained in the introduction, that
the forms of the evaluated
contour integrals are similar, except
for changes of the parameters. One obtains the expressions for the ``universal'' functions
$H^{(1)},H^{(3)},\bar H^{(1)}$ and $\bar H^{(3)}$ by
integrating one specific integral in each case explicitly and
then modifying the definition of the  parameters in these functions to obtain
the other required quantities.

\subsection{Four real zeros of $Q^2(\eta)$ or $ \tilde Q^2(\xi)$ }
\subsubsection{The quantities $\alpha$, $\beta$ and $\bar K$ pertaining
to the $\eta$-equation: Subbarrier case [Fig. 3(a) in Ref. 11]}

Denoting the zeros $\eta_1,\eta_2,\eta_3$ and $\eta_4$ by $a,b,c$ and $d$, 
we write the base function for this case as
\begin{equation}
Q(\eta)=p{{[(\eta - a)(b-\eta)(c-\eta)(d-\eta)]^{\frac {1}{2}}}
\over {1-\eta^2}}.
\end{equation}
Using the transformation on p. 103 in \cite{byrd}, we obtain
\begin{equation}
\eta = {{a-d \nu_1^2sn^2u}\over {1-\nu_1^2sn^2u}},
\end{equation}
the parameter $\nu_1^2$ being defined in (2.4) below. Noting that the loop
$a \rightarrow  b \rightarrow a$, that is 
$\eta_1\rightarrow \eta_2\rightarrow \eta_1$ in the $\eta$-plane, denoted by
$\Gamma_{a,b}$, which represents the contour $\Lambda_{\alpha}$ in Fig. 3(a) of
\cite{ath1}, corresponds to 
$0 \rightarrow K \rightarrow 2K$ in the $u$-plane, one finds that the first-order
contribution to $\alpha$ is

\begin{eqnarray}
\alpha^{(1)}&=&\frac{1}{2}\int_{\Lambda_{\alpha}}Q(\eta)d\eta\nonumber\\
&=&\frac{1}{2}\int_{\Gamma_{a,b}} Q(\eta)d\eta\nonumber\\
&=&{{p}\over {2g}}
\int_0^{2K}\left({{d\eta}\over {du}}\right )^2 {{du}\over {1-\eta^2}}
\nonumber\\
&=&{{2p(a-d)^2\nu_1^4}\over {g(1-a^2)}}\int_0^{2K}
{{sn^2u(1-sn^2u)(1-k^2sn^2u)du}\over {(1-\nu_1^2sn^2u)^2(1-\nu_2^2sn^2u)
(1-\nu_3^2sn^2u)}},
\end{eqnarray}

where
\begin{equation}
\label{par}
\nu_1^2={{a-b}\over {d-b}}, \hskip 10pt\nu_2^2={{1+d}
\over {1+a}}\nu_1^2, \hskip 10pt\nu_3^2={{1-d}
\over {1-a}}\nu_1^2,
\end{equation}
\begin{equation}
\label{par1}
g={{2}\over [(d-b)(c-a)]^{\frac{1}{2}}},\hskip 10pt k^2 ={{(d-c)
(b-a)}\over {(d-b)(c-a)}}.
\end{equation}

Decomposing the integrand in (2.3) into partial fractions and evaluating
the integrals by means of recurrence formulas in \cite{byrd}, we obtain
\begin{equation}
\alpha^{(1)}=H^{(1)}(\nu_1,\nu_2,\nu_3,g,k,C),
\end{equation}
where $H^{(1)}$ is the ``universal'' function defined in eqs.(3.16a,b), (3.17a-c)
 and (3.18) of 
\cite{ath2} but with $\tilde C$ replaced by $C$ and the parameters $\nu_i$, $i=1,2,3$, 
given by (\ref{par}), and $g$ and $k^2$ given by (\ref{par1}). The evaluation of
the quantity $\alpha^{(1)}$ in the present case is thus similar to the
evaluation of the quantity $\tilde L^{(1)}$ in sec. 3.1.2
of \cite{ath2}.

Proceeding in a similar way, the third-order contribution to $\alpha$ is found to be
\begin{eqnarray}
\alpha^{(3)}&=&{{1}\over {2}}\int_{\Lambda_\alpha} \left (-C+{{1}\over {1-\eta^2}}\right ){{d\eta}
\over {2Q(\eta)(1-\eta^2)}}-{{1}\over {16}}\int_{\Lambda_\alpha} Q^{-3}(\eta)\left ({{dQ}\over
{d\eta}}\right)^2d\eta\nonumber\\
&=&H^{(3)}(\nu_1,\nu_2,\nu_3,g,k,C),
\end{eqnarray}
where $H^{(3)}$ is the ``universal'' function given in eqs.(3.16c,d), (3.19a-d) 
and (3.20a-d) of \cite{ath2}
but with $\tilde C$ replaced by $-C$ and the parameters $\nu_i$, $i=1,2,3$, 
given by (\ref{par}), and $g$ and $k^2$ given by (\ref{par1}).

In a similar manner as above we proceed for the calculation of the first- and
third-order contributions to $\beta$. The loop
$d \rightarrow c \rightarrow d$, that is $\eta_4\rightarrow\eta_3\rightarrow\eta_4$
in the $\eta$-plane, denoted by $\Gamma_{d,c}$ and represented by  the contour $\Lambda_{\beta}$ in
Fig. 3(a) of \cite{ath1},
corresponds to $0 \rightarrow K \rightarrow 2K$ in the $u$-plane, and
hence the first- and third-order contributions to $\beta$ are
\begin{subequations}
\begin{eqnarray}
\beta^{(1)}&=&{{1}\over {2}}\int_{\Lambda_{\beta}}Q(\eta)d\eta\nonumber\\
&=&{{1}\over {2}}\int_{\Gamma_{d,c}}Q(\eta)d\eta\nonumber\\
&=&H^{(1)}(\nu_1,\nu_2,\nu_3,g,k,C),
\end{eqnarray}
and
\begin{equation}\beta^{(3)}=H^{(3)}(\nu_1,\nu_2,\nu_3,g,k,C),
\end{equation}
\end{subequations}
where now
\begin{equation}
\nu_1^2={{d-c}\over {d-b}},\hskip 10pt\nu_2^2={{1+b}\over
{1+c}}\nu_1^2,\hskip 10pt\nu_3^2={{1-b}\over {1-c}}\nu_1^2,
\end{equation}
and $g$ and $k^2$ are given by (\ref{par1}).

Similarly one obtains
\begin{subequations}
\begin{equation}
\pi\bar K_0=H^{(1)}(\nu_1,\nu_2,\nu_3,g,k,C),
\end{equation}
\begin{equation}
\pi\bar K_2=H^{(3)}(\nu_1,\nu_2,\nu_3,g,k,C),
\end{equation}
\end{subequations}
where now
\begin{equation}
\nu_1^2={{c-b}\over {d-b}},\hskip 10pt\nu_2^2={{1+d}\over
{1+c}}\nu_1^2,\hskip 10pt\nu_3^2={{1-d}\over {1-c}}\nu_1^2,
\end{equation}
and $g$ is the same as given in (\ref{par1}) and $k^2$ is now given by
\begin{equation}
k^2=\frac{(c-b)(d-a)}{(d-b)(c-a)}.
\end{equation}

According to eq. (3.18a) in \cite{ath1} the integrals $\alpha '$ and $\beta '$ for the contours $\Lambda_{\alpha'}$
and $\Lambda_{\beta '}$ in Fig. 3(a) in \cite{ath1} are obtained from the
formulas $\alpha '=\alpha+\frac{\Lambda\pi}{2}$ and $\beta '=\beta+\frac{
\Lambda\pi}{2}$.

\subsubsection {The quantities $\tilde L$ and $\tilde L'$ pertaining to the 
$\xi$-equation [Fig. 1 in Ref. 11] }

The formulas for the first- and third-order contributions
 to $\tilde L$ and $\tilde L'$ are derived and presented in subsection
3.1.2 of \cite{ath2}, and they remain unchanged in the present case.

\subsection{Two real and two complex conjugate
zeros of $Q^2(\eta)$ or $\tilde Q^2(\xi)$}
When there are two real and two complex conjugate zeros of $Q^2(\eta)$,
the situation of either Fig. 4(a) or Fig. 2 in \cite{ath1} may occur.
The latter situation has, however, so far not appeared in our applications,
and therefore we disregard it in our treatment below of the $\eta$-equation.

\subsubsection{The quantities $\alpha, \beta$, $\bar K$, $L$ and $L'$ pertaining
to the $\eta$-equation: Superbarrier case [Fig. 4(a) or Fig. 2 in Ref. 11]}

Denoting the real zeros of $Q^2(\eta)$ by $\eta_1=a$ and $\eta_4 =d$ and the 
complex conjugate zeros $\eta_2$ and $\eta_3$ by $c$ and $c^*$, we have the base function

\begin{equation}
Q(\eta)=p{{[(a-\eta)(\eta-d)(\eta-c)(\eta-c^*)]^{\frac{1}{2}}}\over {1-\eta^2}}.
\end{equation}
Defining
\begin{equation}
c=b_1-ia_1,\hskip 10pt c^*=b_1+ia_1,
\end{equation}
\begin{subequations}
\begin{equation}
A=[(a-b_1)^2+a_1^2]^{\frac{1}{2}},
\end{equation}
\begin{equation}
B=[(d-b_1)^2+a_1^2]^{\frac{1}{2}},
\end{equation}
\end{subequations}
and using the transformation on p. 133 in \cite{byrd}, we get

\begin{equation}
\label{e1}
\eta ={{aB+dA+(dA-aB)cnu}\over {A+B+(A-B)cnu}}.
\end{equation}

Here we exploit the fact that the Jacobian elliptic functions are doubly 
periodic, one of the periods being complex. Thus the loop $d \rightarrow a_1+ib_1
\rightarrow d$, 
that is $\eta_4\rightarrow \eta_3\rightarrow \eta_4$ in the $\eta$-plane, denoted by $\Gamma_{d,c^*}$, for the contour
$\Lambda_\beta$ in Fig. 4(a) of \cite{ath1}, corresponds in the $u$-plane to the path
$0 \rightarrow K+iK' \rightarrow 2K+2iK'$, where $K$ and $K'$ are complete
elliptic integrals of the modulus $k$, given in (\ref{k1k}) below, and of the 
complementary modulus $k'=\sqrt{1-k^2}$,
respectively. Making
use of the transformation 
(\ref{e1}), we obtain for the integral in the first-order expression for $\beta$
\begin{eqnarray}
\label{sup1}
\frac{1}{2}\int_{\Lambda_\beta}Q(\eta)d\eta&=&\frac{1}{2}\int_{\Gamma_{d,c^*}}Q(\eta)d\eta\nonumber\\
&=&-{{p(\nu_1-\nu_2)(\nu_1-\nu_3)}\over {2g}}\int _0^{2K+2iK'}
{{sn^2udn^2udu}\over {(1+\nu_1 cnu)^2(1+\nu_2cnu)(1+\nu_3cnu)}},
\end{eqnarray}
where
\begin{subequations}
\label{aa}
\begin{equation}
\nu_1={{A-B}\over {A+B}},
\end{equation}
\begin{equation}
\nu_2={{(1+d)A-(1+a)B}\over {(1+d)A+(1+a)B}},
\end{equation}
\begin{equation}\nu_3={{(1-d)A-(1-a)B}\over {(1-d)A+(1-a)B}},
\end{equation}
\begin{equation}
g={{1}\over {\sqrt{AB}}}, 
\end{equation}
\begin{equation}
k^2={{(a-d)^2-(A-B)^2}\over {4AB}}.
\label{k1k}
\end{equation}
\end{subequations}
Similarly one can treat the corresponding integrals in the first-order
expressions for $\alpha, K$ and $\tilde L$. When one then evaluates the integrals
containing the elliptic functions [cf. (2.17)], one finds that all these
integrals can be expressed in terms of a ``universal'' function 
$\bar H^{(1)}$ that is given by
\begin{eqnarray}
\label{sup}
\bar H^{(1)}(\nu_1,\nu_2,\nu_3,g,k,C)&=&-\frac{p}{g}\left\{\frac{1}{\nu_1^2}\left [\left (
2k^2+{{\nu_1^2}\over {1-\nu_1^2}}\right )\Pi\left ({{\nu_1^2}\over 
{\nu_1^2-1}},k \right ) + (\nu_1^2-2k^2)K(k)
\right .\right.\nonumber\\
&&\left.-2\nu_1^2E(k)+ \nu_1k(2j+1)\pi-{{\nu_1^2(1-2k^2)+2k^2}\over 
{[(1-\nu_1^2)(k^2+\nu_1^2k^{'2})]^{\frac{1}{2}}}}
\frac{\nu_1}{2}j\pi\right ]+\sum_{i=1}^3 \bar C_i\bar S_i\nonumber\\
&&\left.+i\left [ \left (2k^2+{{\nu_1^2}\over {1-\nu_1^2}}\right )
\Pi\left({{1}\over{1-\nu_1^2}},k'\right ) -2k^2K(k')+2E(k')\right] 
\right \},
\end{eqnarray}
where $\bar C_1,\bar C_2$ and $\bar C_3$ are given by eqs. (3.37a-c) in
\cite{ath2}, that is,
\begin{subequations}
\begin{equation}
\bar C_1= {{\nu_1^2(2\nu_3\nu_2-\nu_1\nu_3-\nu_1\nu_2)}
\over {(\nu_1-\nu_2)(\nu_1-\nu_3)}},
\end{equation}
\begin{equation}
\bar C_2={{(\nu_1-\nu_3)\nu_2^3}\over {(\nu_1-\nu_2)(\nu_2-\nu_3)}},
\end{equation}
\begin{equation}
\bar C_3={{(\nu_1-\nu_2)\nu_3^3}\over {(\nu_1-\nu_3)(\nu_3-\nu_2)}},
\end{equation}
\end{subequations}
and
\begin{eqnarray}
\bar S_i&=&{{1}\over {\nu_i^4}}\left[k^2(1-\nu_i^2)K(k)+\nu_i^2E(k)
-(k^2+\nu_i^2k^{'2})\Pi\left({{\nu_i^2}
\over {\nu_i^2-1}},k\right )\right]\nonumber\\
&&+{{1}\over {2k\nu_i^3}}\left [\left (k^2(\nu_i^2-1)-{{\nu_i^2}\over {2}}\right )(2j+1)\pi
+k[(1-\nu_i^2)(k^2+k^{'2}\nu_i^2)]^{\frac{1}{2}}j\pi\right ]\nonumber\\ 
&&+{{i}\over {\nu_i^2}}\left \{-(k^2+\nu_i^2k^{'2})\left [\Pi\left({{1}\over
{1-\nu_i^2}},k'\right ) -K(k')\right ]-E(k')\right\}, i=1,2,3,
\end{eqnarray}
$j$ being an integer $-1,0 $ or $+1$, depending upon whether the quantity
$\beta^{(1)}$, $L^{(1)}$ or $\alpha^{(1)}$, respectively, is evaluated, and
originating from  terms $tan^{-1}(sdu)$ and $cos^{-1}(dn u)$ while applying the limits
of integration. Note that the quantity $\tilde L^{(1)}$ given
by eq.(3.36a) in \cite{ath2}, is just $-2Re\bar H^{(1)}$ with $j=0$. For
$K$, which is expressed in terms of the imaginary part of $\bar
H^{(1)}$, the value of $j$, which appears only in the real part of
$\bar H^{(1)}$, does not matter.

Similarly we also introduce the ``universal'' function $\bar H^{(3)}$ given by

\begin{eqnarray}
\label{sups1}
\bar H^{(3)}(\nu_1,\nu_2,\nu_3,g,k,C)& =&{{g}\over {16p}}\left [\left 
(-4C+{{\nu_1^2(\nu_2-\nu_3)^2}
\over {\nu_2 \nu_3(\nu_1-\nu_3)(\nu_2-\nu_1)}}\right ) K(k)\right.\nonumber\\ 
&&+\left. {{1}\over {(\nu_1-\nu_3)(\nu_1-\nu_2)}} \left(X K(k)+Y E(k)
+\frac{2Zj}{k}\pi\right ) \right ]\nonumber\\
&&+i{{g}\over{16p}}\left [\left (-4C+{{\nu_1^2(\nu_2-\nu_3)^2}
\over {\nu_2 \nu_3(\nu_1-\nu_3)(\nu_2-\nu_1)}}\right ) K(k')\right.\nonumber\\
&&+\left. {{1}\over {(\nu_1-\nu_3)(\nu_1-\nu_2)}} \left\{X K(k') +Y[K(k')-E(k')]
\right\}\right],
\end{eqnarray}
where $C$ is the parameter in the square of the base function $Q(\eta)$ in eq. (3.2b)
of \cite{ath1}, and 
\begin{subequations}
\begin{eqnarray}
X&=&-{{1+4k^2}\over {3}}+{{3+4k^{'2}}\over {3}}(\nu_1^2+2\nu_1
\nu_2+2\nu_1\nu_3+\nu_2\nu_3)-{{k^{'2}}\over {3k^2}}(17-4k^2)\nu_1^2\nu_2\nu_3\nonumber\\
&&-2\nu_1(\nu_1+\nu_2+\nu_3)+2\nu_2\nu_3+\nu_1^2\left(\frac{\nu_2}{\nu_3}
+\frac{\nu_3}{\nu_2}\right),
\end{eqnarray}

\begin{eqnarray}
Y&=&{{1}\over {3k^{'2}}}(1+8k^2-8k^4)+{{4}\over {3}}(2k^2-1)(\nu_1^2+2\nu_1
\nu_2+2\nu_1\nu_3+\nu_2\nu_3) \nonumber\\
&&+{{\nu_1^2\nu_2\nu_3}\over
{3k^2}}(17-8k^2+8k^4),
\end{eqnarray}

\begin{equation}
Z=2\nu_1(\nu_1\nu_2+\nu_1\nu_3+4\nu_2\nu_3),
\end{equation}
\end{subequations}
where the parameters $\nu_1,\nu_2,\nu_3,g$ and $k^2$ are defined
in (\ref{aa}a-e). 

Then the first- and third-order contributions to the quantities $\alpha, \beta$
and $K$ $(=\pi\bar K)$ are 

\begin{equation}
\alpha^{(1)}= \mbox{Re} \bar H^{(1)} \hskip 10pt with \hskip 5pt j=1,
\end{equation}
\begin{equation}
\beta^{(1)}=\mbox{Re} \bar H^{(1)} \hskip 10pt with \hskip 5pt j=-1,
\end{equation}
\begin{equation}
\pi\bar K_0 = -2 \mbox{Im} \bar H^{(1)}, 
\end{equation}
\begin{equation}
\alpha^{(3)}= \mbox{Re} \bar H^{(3)}\hskip 10pt with \hskip 5pt j=1,
\end{equation}
\begin{equation}
\beta^{(3)}=\mbox{Re} \bar H^{(3)} \hskip 10pt with \hskip 5pt j=-1,
\end{equation}
\begin{equation}
\pi\bar K_2=-2 \mbox{Im} \bar H^{(3)},
\end{equation}
where $\nu_1$, $\nu_2$, $\nu_3$, $g$ and $k^2$ are still defined by
(\ref{aa}a-e). Since $j$ appears only in the real part of $\bar
H^{(2n+1)}$, we need not specify a value of $j$ in (2.26) and (2.29).

The integrals $\alpha'$ and $\beta'$ associated with the contours 
$\Lambda_{\alpha'}$ and $\Lambda_{\beta'}$ in Fig. 4(a) in \cite{ath1} are
obtained from $\alpha$ and $\beta$ by means of the relations (3.18a) in
\cite{ath1}, that is, $\alpha'=\alpha+\frac{\Lambda\pi}{2}$ and 
$\beta'=\beta+\frac{\Lambda\pi}{2}$. The integrals $L$ and $L'$ associated
with the contours $\Lambda_L$ and $\Lambda_{L'}$ in Fig. 4(a) in \cite{ath1} 
can be obtained from $\alpha$ and $\beta$ by means of the formulas
$L=\alpha+\beta$ and $L'=L+|m|$. 

\subsubsection{The quantities $\tilde L$ and $\tilde L'$
 pertaining to the $\xi$-equation [Fig. 1(a) in Ref. 11]}

The formulas for the first- and third-order contributions to $\tilde
L$ and $\tilde L'$ remain the same as the ones presented in subsection
3.2.2 of \cite{ath2}.

\section{Case $\Lambda =0$}
\subsection{Two real zeros of $Q^2(\eta )$ or $\tilde Q^2(\xi )$}

One should be able to obtain the formulas pertaining to the case $\Lambda=0$
by considering the limits of the ``universal'' functions $H^{(1)}$ and
$H^{(3)}$ when $a \rightarrow -1 \mbox{ and } d 
\rightarrow +1$ in the case of the $\eta$-equation. These specialization
procedures are, however, much more cumbersome than the direct calculation
of the quantities in question, and so we do not carry them out here. Instead
we shall evaluate these quantities directly, and therefore no
``universal'' functions will appear in subsection 3.1.

\subsubsection{The quantities $\alpha$, $\beta $ and $\bar K$ pertaining to the  
$\eta $-equation: Subbarrier case [Fig. 3(b) in Ref. 11]}
 
We denote the two real zeros of $Q^2(\eta)$ by $\eta_2=b$ and $\eta_3=c$
and use transformations on p. 103, p. 120 and p. 112 in \cite{byrd} for
calculating the first- and third-order contributions to
$\alpha$, $\beta$ and $K$, respectively.
 Here the base function reads as

\begin{equation}
Q(\eta ) = p\left [{{(\eta  -b)(c-\eta)}\over {1-\eta ^2}}\right]^{\frac {1}{2}}.
\end{equation}

\noindent The first- and third-order contributions to $\alpha$ are
\begin{subequations}
\begin{equation}
\alpha^{(1)} = {{2p}\over {g\nu^2}}[\nu^2E(k)-(k^2+\nu^2)K(k)+(k^2-\nu^4)
\Pi(\nu^2,k)],
\end{equation}  
and
\begin{eqnarray}
\alpha^{(3)}&=&-{{gCK(k)}\over{p}}+{{g}\over{4p\nu^2}}\left[\left(1-2\nu^2+{{\nu^4}
\over{k^2}}\right )K(k)-\left(1+{{\nu^4}\over{k^2}}\right)E(k)\right]\nonumber\\
&&-{{g}\over{32p\nu^2}}\left \{{{1}\over{3k^{'2}}}\left[\frac{1}{k^2}(4\nu^4
-6\nu^2)+5\nu^4+4\nu^2+8-k^2(4+3\nu^4)-{{6\nu^4}\over{k^4}}\right]K(k)\right.\nonumber\\
&&\left.+{{1}\over{3k^{'4}}}\left[-8k^4+(2+8\nu^2+4\nu^4)k^2-8+2\nu^4+\nu^2
-{{1}\over{k^2}}(2\nu^4+12\nu^2)+{{12\nu^4}\over{k^4}}\right]E(k)\right\},
\end{eqnarray}
\end{subequations}
where 
\begin{equation}
\label{g}
\nu^2={{c+1}\over {c-1}},\hskip 10pt k^2={{(1-b)(1+c)}\over
{(1+b)(1-c)}}, \hskip 10ptg={{2}\over {[(1-c)(1+b)]^{\frac{1}{2}}}}.
\end{equation}
\vskip 5pt

Similarly the first- and third-order contributions to $\beta$ are
\begin{subequations}
\begin{equation}
\beta^{(1)}={{2p}\over {g\nu^2}}[\nu^2E(k)+(k^2-\nu^2)K(k)+(\nu^4
-k^2)\Pi(\nu^2,k)],
\end{equation}
and
\begin{eqnarray}
\beta^{(3)}&=&-{{gCK(k)}\over {2p}}+{{g}\over {4p\nu^2(1-c)}}\left \{
\left(1-{{\nu^4}\over {k^2}}\right )K(k)
+\left [k^2(2k^2-1-2\nu^2)+\nu^4\right ]
{{E(k)}\over {k^2k^{'2}}} \right \}\nonumber\\
&&-{{g}\over {16(1-c)p\nu^2}} \left \{{{4}\over {3k^2}}\left [ (2+\nu^4+2\nu^2)
k^2-3k^4-2\nu^4\right ]K(k)\right.\nonumber\\
&&\left.+{{8}\over {3k^{'2}}}
\left (-(\nu^4+\nu^2-1)
+k^2(\nu^4-\nu^2+2)+{{\nu^4}\over {k^2}}\right )E(k)\right \} 
\end{eqnarray}
\end{subequations}
with

\begin{equation}
\nu^2={{1-b}\over {1-c}}, 
\end{equation}
and $k^2$ and $g$ defined in (\ref{g}). 
 
The first- and third-order contributions to $ K (=\pi\bar K)$ are
\begin{subequations}
\begin{equation}\pi \bar K_0 =  {{2p}\over {gk^2}}
[(k^2-\nu^2)K(k)-\nu^2E(k)-(\nu^4-2\nu^2+k^2)\Pi(\nu^2,k)],
\end{equation}
and
\begin{eqnarray}
\pi\bar K_2&=&-{{CgK(k)}\over{2p}}+{{g}\over{2(c^2-1)p}}\left[
{{2\nu^2}\over{k^4}}(k^2-\nu^2)K(k)+{{1}\over{k^4k^{'2}}}\{k^2(k^2-\nu^4-2\nu^2)
+2\nu^4\}E(k)\right]\nonumber\\
&&-{{g}\over{24\nu^4p(c-1)}}\left\{(-2+k^2+\nu^2-\nu^4)K(k)\right.\nonumber\\
&&\left.+2[k^4-(1+\nu^2+\nu^4)k^2
-\nu^2+2\nu^4+1]E(k)\right\},
\end{eqnarray}
\end{subequations}
with
\begin{equation}
\nu^2={{b-c}\over {b+1}}, \hskip 10ptk^2={{2(b-c)}\over {(1+b)(1-c)}},
\end{equation}
and $g$ defined in (\ref{g}).

\subsubsection {The quantities  $\tilde L$ and $\tilde L'$
 pertaining to the $\xi$-equation [Fig. 1 in Ref. 11]}
The formulas for the first- and third-order contributions 
to $\tilde L$ and $\tilde L'$ are the same as those derived and presented 
in subsection 4.1.2 of \cite{ath2}.

\subsection {Two complex conjugate zeros of $Q^2(\eta)$}
The situation of two complex conjugate transition zeros can occur only
for $Q^2(\eta)$ but not for $\tilde Q^2(\xi)$.
\subsubsection{The quantities
$\alpha, \beta$ and $\bar K$ pertaining to the $\eta $ equation: 
Superbarrier case \\ \mbox{ }[Fig. 4(b) in Ref. 11]}
Specializing the general formulas (\ref{sup}) and (\ref{sups1}) 
by putting $a=-1,d=+1$, we obtain
\begin{subequations}
\begin{eqnarray}
\bar H^{(1)}&=&{{p}\over {g}}\left \{E(k)-K(k)+{{1}\over {1-\nu^2}}\left[\Pi\left({{\nu^2}
\over {\nu^2-1}},k\right )-
 \left({{\nu^2(1-\nu^2)}\over {k^2+(1-k^2)\nu^2}}\right)^\frac{1}{2}(2j+1)\pi
\right]\right\}\nonumber\\
&&+i{{p}\over {g}}\left [K(k')-E(k')+{{\nu^2}\over {1-\nu^2}}\Pi\left ( {{1}\over {1-\nu^2}}
,k'\right ) \right ],
\end{eqnarray}
and
\begin{eqnarray}
\bar H^{(3)}&=&\frac{g}{16p}\left\{\left(-4C+\frac{4\nu^2}{\nu^2-1}\right)K(k)
+\frac{1}{\nu^2-1}\left(\bar X K(k)+\bar Y E(k)-\frac{16\nu j}{k}\pi\right)\right.\nonumber\\
&&\left.+i\left[\left(-4C+\frac{4\nu^2}{\nu^2-1}\right)K(k')
+\frac{1}{\nu^2-1}\left(\bar X K(k')+\bar Y [K(k')-E(k')]\right)\right]\right\},
\end{eqnarray}
\end{subequations}
where
\begin{equation}
\nu={{A-B}\over {A+B}}, \hskip 10pt g={{1}\over {\sqrt{AB}}},\hskip 10pt k^2={{4-(A-B)^2}\over
{4AB}}
\end{equation}
with

\begin{equation}
A=[(1-b_1)^2+a_1^2]^\frac{1}{2}, \hskip
10ptB=[(1+b_1)^2+a_1^2)]^\frac{1}{2}
\end{equation}
and
\begin{subequations}
\begin{equation}
\bar X=-\frac{1}{3k^2}[k^2(14+9\nu^2)+17\nu^2],
\end{equation}
\begin{equation}
\bar Y=\frac{1}{3k^2k^{'2}}[-4k^4(1+\nu^2)+k^2(5+21\nu^2)-17\nu^2].
\end{equation}
\end{subequations}
We now have


\begin{subequations}
\begin{equation}
\alpha^{(1)}=\mbox{Re} \bar H^{(1)} \mbox{ with} \hskip 5pt j=1,
\end{equation}
\begin{equation}
\alpha^{(3)}=\mbox{Re} \bar H^{(3)} \mbox{ with} \hskip 5pt j=1,
\end{equation}
\end{subequations}
\begin{subequations}
\begin{equation}
\beta^{(1)}= \mbox{Re} \bar H^{(1)} \mbox{ with} \hskip 5pt j=-1, 
\end{equation}
\begin{equation}
\beta^{(3)} = \mbox{Re} \bar H^{(3)} \mbox{ with} \hskip 5pt j=-1,
\end{equation}
\end{subequations}
\begin{subequations}
\begin{equation}
\pi \bar K_0 =-2\mbox{Im} \bar H^{(1)},
\end{equation}
\begin{equation}
\pi\bar K_2 = -2\mbox{Im} \bar H^{(3)}.
\end{equation}
\end{subequations}

The integral $L'$ associated with the contour $\Lambda_{L'}$ in Fig.
4(b) in \cite{ath1} is obtained from the formula $L'=\alpha +\beta$.
\section {Numerical illustration of the accuracy of the quantization
conditions}

For the numerical illustration of the asymmetric case we have chosen $Z_1=1$ and considered
three different values of $Z_2$, {\it viz}. $Z_2=2,\hskip 5pt 5 \hskip 5pt\mbox{and}\hskip 5pt 8$.
 The corresponding physical systems  are the ions 
$peHe^{2+}$,\hskip 5pt $peB^{5+}$ and $peO^{8+}$, respectively, where p is a proton and e is an electron.
For each one of these systems we have calculated the eigenvalue $p$ and the reduced separation
constant $A'$ for two different $\sigma$-states and various values of $r_{12}$.

For the ion $peHe^{2+}$ we have calculated the eigenvalue $p$ and the reduced separation
constant $A'$ for the $1s\sigma$ and $2p\sigma$ states and various values of
$r_{12}$, with appropriate quantization conditions and parameters.
 The quantization
conditions  in \cite{ath1} for the $1s\sigma$ state are
(3.5a) with $\tilde s=0$ [Fig. 1(a) in Ref. 11] and (3.9) with $s=m=0$ 
[Fig. 4(b) in Ref. 11] when $r_{12}$ is
sufficiently small, but (3.5b) with $\tilde s =m=0$ [Fig. 1(b) in Ref. 11]  and (3.25b) with
$s_\beta = m =0$ [Fig. 3(b) in Ref. 11] when $r_{12}$ is sufficiently large. The quantization
conditions in \cite{ath1} for the $2p\sigma$ state are (3.5a) with $\tilde s=0$
[Fig. 1(a) in Ref. 11] and (3.9) with $s=1$ and $m=0$ [Fig. 4(b) in Ref. 11]  when $r_{12}$ is sufficiently small, but (3.5b)
with $\tilde s = m =0$ [Fig. 1(b) in Ref. 11]  and (3.25a) with $s_\alpha = m =0$
[Fig. 3(b) in Ref. 11] when $r_{12}$ is
sufficiently large.

For the ion $peB^{5+}$ we have computed the eigenvalue $p$ and the reduced separation constant
$A'$ for the $1s\sigma$ and
the $3s\sigma$ states and various values of $r_{12}$, 
with appropriate quantization conditions and parameters.
 The quantization
conditions  in \cite{ath1} for the $1s\sigma$ state are
(3.5a) with $\tilde s=0$ [Fig. 1(a) in Ref. 11]  and (3.9)  with $s=m=0$  
[Fig. 4(b) in Ref. 11] when $r_{12}$ is
sufficiently small, but (3.5b) with $\tilde s =m=0$  [Fig. 1(b) in Ref. 11] 
and (3.25b) with
$s_\beta = m =0$  [Fig. 3(b) in Ref. 11] when $r_{12}$ is sufficiently large. The quantization
conditions in \cite{ath1} for the $3s\sigma$ state are (3.5a) with $\tilde s=2$
 [Fig. 1(a) in Ref. 11] and (3.9)  with $s=m=0$  [Fig. 4(b) in Ref. 11]  when $r_{12}$ is sufficiently small, but (3.5b)
with $\tilde s = m =0$  [Fig. 1(b) in Ref. 11] and (3.25b) with $s_\beta = m =0$  
[Fig. 3(b) in Ref. 11] 
  when $r_{12}$ is
sufficiently large.

For the ion $peO^{8+}$ we have calculated the eigenvalue $p$ and the reduced separation constant
$A'$ for the $1s\sigma$ and
the $4d\sigma$ states and various values of $r_{12}$, 
with appropriate quantization conditions and parameters.
 The quantization
conditions  in \cite{ath1} for the $1s\sigma$ state are
(3.5a) with $\tilde s=0$ [Fig. 1(a) in Ref. 11] and (3.9)  with $s=m=0$ 
[Fig. 4(b) in Ref. 11] when $r_{12}$ is
sufficiently small, but (3.5b) with $\tilde s =m=0$  [Fig. 1(b) in Ref. 11] and (3.25b) with
$s_\beta = m =0$  [Fig. 3(b) in Ref. 11] when $r_{12}$ is sufficiently large. The quantization
conditions in \cite{ath1} for the $4d\sigma$ state are (3.5a) with $\tilde s=1$
 [Fig. 1(a) in Ref. 11] and (3.9)  with $s=2$ and $m=0$ [Fig. 4(b) in Ref. 11] when $r_{12}$ is sufficiently small, but (3.5b)
with $\tilde s = m =0$ [Fig. 1(b) in Ref. 11] and (3.25b) with $s_\beta=2$ and $m =0$ 
 [Fig. 3(b) in Ref. 11] when $r_{12}$ is
sufficiently large.

In the calculations for the above mentioned three ions we used  the quantization
conditions expressed in terms of complete elliptic integrals, obtained in the present paper,
that correspond to the above mentioned quantization conditions in \cite{ath1}.
In  
subsection 4.1 we use the same procedure as in subsection 5.1 of \cite{ath2}
for optimizing the accuracy of the results obtainable in the first- and third-order
approximations. Thus we determine $C$ and $\tilde C$ as  functions of $r_{12}$
such that the first- and third-order quantization conditions give the same results,
and with these values of $C$ and $\tilde C$ we calculate $p$ and $A'$.
In subsection 4.2 we determine $C$ and $\tilde C$ such that the
phase-integral quantization conditions give the numerically exact values of
$p$ and $A'$ obtained by Winter {\it et al.} \cite{win} for the ion $peHe^{2+}$  and by 
Ponomarev and Puzynina\cite{pon} for the ions $peB^{5+}$ and $peO^{8+}$.
It is seen that the values of $C$ and $\tilde C$ obtained in subsection
4.1 are in qualitative agreement with those obtained in subsection
4.2.

\subsection{Determination of $C(r_{12})$ and $\tilde C(r_{12})$ such that
the first- and third-order quantization conditions give the same values of
$p$ and $A'$}

We have determined the values of $C$ and $\tilde C$ for several values of
$r_{12}$ such that the first- and third-order quantization conditions give
the same value of $p$ as well as of $A'$ . These values are tabulated 
 and compared with the numerically exact results obtained by
Winter {\it et al.} \cite{win} for the ion $peHe^{2+}$ and by Ponomarev and 
Puzynina\cite{pon} for the ions $peB^{5+}$ and $peO^{8+}$. 
In Table I and Table II we give the results for
the system $peHe^{2+}$. Table III and Table IV present the results
for the ion $peB^{5+}$. For the ion $peO^{8+}$ the results are tabulated
in Table V and Table VI. The results  in Tables I - VI are presented
graphically in Figs. 1 - 6.
 
\subsection{Determination of $C(r_{12})$ and $\tilde C(r_{12})$ such that
the phase-integral quantization conditions reproduce numerically exact
values of $p$ and $A'$}

By determining $C$ and $\tilde C$ for each value of $r_{12}$ such that the first-order
quantization conditions
reproduce the numerically exact values of $p$ and $A'$ obtained by Winter {\it et al.} \cite{win}
for the $1s\sigma$ and $2p\sigma$ states of the ion $peHe^{2+}$ and by Ponomarev
and Puzynina\cite{pon} for the previously mentioned two states of the ion 
$peB^{5+}$ and two states of the ion $peO^{8+}$ 
we have obtained the values of $C$ and $\tilde C$
presented in Tables VII - XII. The numerical results in these tables are presented
graphically in Figs. 7 - 12.

\vskip 10pt
\acknowledgments
We are much indebted to Professor T. G. Winter for placing the unpublished
numerical material mentioned on p. 288-289 in \cite{win} at our
disposal. Also the authors are extremely grateful
to Professor Per Olof Fr\"oman for very critical reading of the manuscript and making
numerous comments which resulted in a much improved presentation.
The work of M.L. forms part of a Department of Science and Technology, Government
of India, research project. Support from the Swedish Natural Science Research Council
for M. Lakshmanan's visits to Uppsala is gratefully acknowledged.

\begin{table}

\caption{ For the state $1s\sigma$ of the ion $peHe^{2+}$ $(Z_1=1,
Z_2=2)$ the values of $C$ and $\tilde C$ have been obtained from the
requirement that the first- and third-order phase-integral results
coincide for $p$ as well as for $A'$. With the use of these values of
$C$ and $\tilde C$ the  values of $p$ and $A'$ have then been obtained
from the quantization conditions that are appropriate depending on
whether $r_{12}$ is sufficiently small or sufficiently large. The
numerically exact values (accurate to all digits quoted) calculated by
Winter, Duncan and Lane [4], and obtained as private communication from
Professor Winter (see p. 288-289 in [4]), are given in the columns
called $p_{WDL}$ and $A'_{WDL}$.  }

\begin{tabular}{ccccccccc}
\hline
$r_{12}$ & $C$ & $\tilde C$ & $p$ & $p_{WDL}$ & $p-p_{WDL}$ & $A'$&$A'_{WDL}$ & $A'-A'_{WDL}$ \\
\hline
\multicolumn{9}{l} {}\\
\multicolumn{9}{l} {Sufficiently small $r_{12}$}\\
\hline
0.2&0.4296362900&0.4986159300&0.2913835957&0.2909534228&0.000430173&-0.0401742480&-0.0495531186&0.00937887\\
\hline
0.4&0.4938610210&0.4991342047&0.5546735631&0.5544040477&0.000269516&-0.16745905&-0.1752443935&0.007785348\\
\hline
0.6&0.5014529650&0.5021936418&0.7949730256&0.7945061056&0.00046692&-0.3404297821&-0.3475381522&0.00710837\\
\hline
0.8&0.5023672195&0.5045291831&1.018520819&1.018366017&-0.000154802&-0.54176553&-0.5473748938&0.00560936\\
\hline
1.0&0.5032793418&0.5061539132&1.231430173&1.231534107&-0.000103934&-0.75725945&-0.7624147481&0.0051553\\
\hline
2.0&0.5059145550&0.5083606740&2.241478757&2.241514227&-0.00003547&-1.870558067&-1.866548007&-0.00401006\\
\hline
\multicolumn{9}{l} {}\\
\multicolumn{9}{l} {Sufficiently large $r_{12}$}\\
\hline
3.0&0.4951651150&0.5099904870&3.241389959&3.241868168&-0.000478209&-2.918225629&-2.914992386&-0.003233243\\
\hline
4.0&0.4904409110&0.5094308280&4.243060447&4.243211413&-0.000150966&-3.938607817&-3.937060587&-0.00154723\\
\hline
5.0&0.4878197130&0.5085272790&5.244268024&5.244326655&-0.000058631&-4.950729976&-4.949835242&-0.000894734\\
\hline
6.0&0.4861586190&0.5076632010&6.245131490&6.245159553&-0.000028063&-5.958842182&-5.958257005&-0.000585177\\
\hline
7.0&0.4850145200&0.5069119800&7.245774185&7.245789653&-0.000015468&-6.964658345&-6.964245417&-0.000412928\\
\hline
8.0&0.4841801400&0.5062737200&8.246269561&8.246278979&-0.000009418&-7.969033663&-7.968726708&-0.000306955\\
\hline
9.0&0.4835456720&0.5057325250&9.246662366&9.246668544&-0.000006178&-8.972444904&-8.972207821&-0.000237083\\
\hline
10.0&0.4830475910&0.5052712520&10.24698113&10.24698542&-0.00000429&-9.975179205&-9.974990618&-0.000188587\\
\hline
11.0&0.4826466130&0.5048751070&11.24724480&11.24724792&-0.00000312&-10.97741991&-10.97726635&-0.00015356\\
\hline
12.0&0.4823171600&0.5045321160&12.24746643&12.24746878&-0.00000235&-11.97928963&-11.97916218&-0.00012745\\
\hline
13.0&0.4820418880&0.5042327580&13.24765526&13.24765709&-0.00000183&-12.98087344&-12.98076598&-0.00010746\\
\hline
14.0&0.4818086000&0.5039695200&14.24781804&14.24781949&-0.00000145&-13.98223227&-13.98214045&-0.00009182\\
\hline
15.0&0.4818086000&0.5037364370&15.24795978&15.24796097&-0.00000119&-14.98341087&-14.983331513&-0.00007936\\
\hline
\end{tabular}
\end{table}

\begin{table}
\caption{
For the state $2p\sigma$ of the ion $peHe^{2+}$ $(Z_1=1, Z_2=2)$ the values of $C$ and $\tilde C$ have been obtained
from the requirement that the first- and third-order phase-integral results
coincide for $p$ as well as for $A'$. With the use of these values of $C$ and $\tilde C$
the  values of $p$ and $A'$ have then been obtained from the 
quantization conditions that are appropriate depending
on whether $r_{12}$ is sufficiently small or sufficiently large. The numerically exact values
(accurate to all digits quoted) calculated by Winter, Duncan and Lane [4], and obtained as private
communication from Professor Winter (see p. 288-289 in [4]), are given in the columns called $p_{WDL}$
and $A'_{WDL}$.
}
\begin{tabular}{ccccccccc}
\hline
$r_{12}$ & $C$ & $\tilde C$ & $p$ & $p_{WDL}$ & $p-p_{WDL}$ & $A'$&$A'_{WDL}$ & $A'-A'_{WDL}$ \\
\hline
\multicolumn{9}{l} {}\\
\multicolumn{9}{l} {Sufficiently small $r_{12}$}\\
\hline
0.2&0.5294831952&0.5029173619&0.1517930283&0.1507994078&0.000993621&-2.009318372&-2.013114367&0.003795995\\
\hline
0.4&0.5481736492&0.5102837463&0.3028910382&0.3062680406&-0.003377002&-2.049183021&-2.053824124&0.004641103\\
\hline
0.6&0.5928374081&0.5279102827&0.468491904&0.4698372597&-0.001345355&-2.120325324&-2.125935288&0.0056099640\\
\hline
0.8&0.6493296518&0.5380281837&0.642918371&0.6416038747&0.001314497&-2.226301818&-2.234147173&0.007845355\\
\hline
1.0&0.6918276523&0.5400183926&0.819207383&0.8180287700&0.001178613&-2.382017327&-2.381560387&-0.00045694\\
\hline
2.0&0.7291847142&0.5410643929&1.641039285&1.640235157&0.000804128&-3.839201840&-3.846791567&0.007589727\\
\hline
\multicolumn{9}{l} {}\\
\multicolumn{9}{l} {Sufficiently large $r_{12}$}\\
\hline
3.0&0.7931298547&0.5426329856&2.304295844&2.303194434&0.00110141&-5.443294571&-5.444185235&0.000890664\\
\hline
4.0&0.8368432961&0.5429173921&2.869430632&2.872046343&-0.002615711&-7.447320938&-7.448941809&0.001620871\\
\hline
5.0&0.9439467950&0.5427393690&3.398416434&3.395848335&0.002568099&-9.524254856&-9.526950457&0.002695601\\
\hline
6.0&0.7710812780&0.5422035320&3.913851362&3.901954918&0.011896444&-11.58670782&-11.61569923&0.02899141\\
\hline
7.0&0.7080781570&0.5407494330&4.425841569&4.404864367&0.020977202&-13.63385383&-13.68696569&0.05311186\\
\hline
8.0&0.6709709560&0.5389585440&4.933037475&4.909225930&0.023811545&-15.67683067&-15.73665989&0.05982922\\
\hline
9.0&0.6449125300&0.5370377600&5.438128294&5.414760984&0.02336731&-17.71314008&-17.77106869&0.05792861\\
\hline
10.0&0.6252417660&0.5351150510&5.942311467&5.920460066&0.021851401&-19.74285483&-19.79629610&0.05344127\\
\hline
11.0&0.6098017850&0.5332640160&6.445985745&6.425795942&0.020189803&-21.76715693&-21.81592389&0.04876696\\
\hline
12.0&0.5973476560&0.5315208600&6.949277104&6.930614764&0.01866234&-23.78726983&-23.83184680&0.04457697\\
\hline
13.0&0.5870841030&0.5298992780&7.452239113&7.434919206&0.017319907&-25.80416211&-25.84512132&0.04095921\\
\hline
14.0&0.5784759070&0.5284005350&7.954909402&7.938760040&0.016149362&-27.81854786&-27.85639810&0.03785024\\
\hline
15.0&0.5711495800&0.5270195800&8.457321437&8.442196146&0.015125291&-29.83094919&-29.86611405&0.03516486\\
\hline
\end{tabular}
\end{table}

\begin{table}[ht!]
\caption{
For the state $1s\sigma$ of the ion $peB^{5+}$ $(Z_1=1, Z_2=5)$ the values of $C$ and $\tilde C$ have been obtained
from the requirement that the first- and third-order phase-integral results
coincide for $p$ as well as for $A'$. With the use of these values of $C$ and $\tilde C$
the  values of $p$ and $A'$ have then been obtained from the 
quantization conditions that are appropriate depending
on whether $r_{12}$ is sufficiently small or sufficiently large. The numerically exact values
obtained by Ponomarev and Puzynina [13] are given in the columns called $p_{PP}$
and $A'_{PP}$.
}
\begin{tabular}{ccccccccc}
\hline
$r_{12}$ & $C$ & $\tilde C$ & $p$ & $p_{PP}$ & $p-p_{PP}$ & $A'$&$A'_{PP}$ & $A'-A'_{PP}$ \\
\hline
\multicolumn{9}{l} {}\\
\multicolumn{9}{l} {Sufficiently small $r_{12}$}\\
\hline
0.2&     0.51832&    0.48329&      0.576736&  0.57180&  0.004936&   -0.102639&    -0.106339&  0.003694\\  
\hline

0.4&     0.51568&    0.48937&      1.093781&  1.09319&  0.000591&   -0.318422&    -0.312917&  0.005505\\
\hline

0.6&    0.51379&   0.49583&     1.597092&1.59754&   -0.000448&  -0.539704&    -0.533666&-0.006038\\
\hline
    
0.8&    0.51057&   0.49961&     2.097826&2.09827&-0.000444&  -0.753875&    -0.749331&-0.004544\\
\hline

1.0&    0.50847&   0.50137&     2.598201&2.59847&  -0.000269&   -0.962648&-0.959583&-0.003065\\
\hline

2.0&    0.50456&   0.50273&     5.099079&5.09906&  +0.000019&    -1.980607&   -1.97997&-0.000637\\
\hline
\multicolumn{9}{l} {}\\
\multicolumn{9}{l} {Sufficiently large $r_{12}$}\\
\hline
3.0&    0.50320&   0.50231&     7.599399&7.59936&+0.000039&   -2.986865&      -2.98666&-0.000205\\
\hline

4.0&    0.50230&  0.50192&    10.099513&10.0995&+0.000013&  -3.990163&       -3.99000&-0.000163\\
\hline

5.0&    0.50081&   0.50164&    12.599680& 12.5996&+0.000080&  -4.991936&      -4.99200&+0.000064\\
\hline

6.0&    0.50090&   0.50141&    15.099774&15.0997& +0.000074&   -5.993175&     -5.99333&+0.000155\\
\hline

7.0&    0.50082&   0.50124&    17.599832&17.5997&+0.000132&  -6.994086&       -6.99428&+0.000194\\
\hline

8.0&    0.50081&   0.50111&    20.099870&20.0998&+0.000070&   -7.994774&     -7.99500&+0.000226\\
\hline

9.0&    0.50080&   0.50100&    22.599897& 22.5998&+0.000097&   -8.995344&     -8.99556&+0.000026\\
\hline

10.0&   0.50080&   0.50090&    25.099916&25.0998&+0.000116&  -9.995783&       -9.99563&-0.000153\\
\hline

11.0&   0.50070&   0.50083&    27.599930&27.5998&+0.000130& -10.996136&      -10.9960&-0.000136\\
\hline

12.0&   0.50070&   0.50080&    30.099941&30.0998&+0.000140& -11.996448&      -11.9964&-0.000048\\
\hline

13.0&   0.50060&   0.50067&    32.599950&32.5998& +0.000150& -12.996740&     -12.9966&-0.000140\\
\hline

14.0&   0.50070&   0.50060&    35.099957&35.0998&+0.000157& -13.996981&      -13.9969&-0.000081\\
\hline

15.0&   0.50060&   0.50060&    37.599962& 37.5999& +0.000062&-14.997142&     -14.9971&-0.000042\\
\hline
\end{tabular}
\end{table}

\begin{table}
\caption{
For the state $3s\sigma$ of the ion $peB^{5+}$ $(Z_1=1, Z_2=5)$ the values of $C$ and $\tilde C$ have been obtained
from the requirement that the first- and third-order phase-integral results
coincide for $p$ as well as for $A'$. With the use of these values of $C$ and $\tilde C$
the  values of $p$ and $A'$ have then been obtained from the 
quantization conditions that are appropriate depending
on whether $r_{12}$ is sufficiently small or sufficiently large. The numerically exact values
obtained by Ponomarev and Puzynina [13] are given in the columns called $p_{PP}$
and $A'_{PP}$.
}
\begin{tabular}{ccccccccc}
\hline
$r_{12}$ & $C$ & $\tilde C$ & $p$ & $p_{PP}$ & $p-p_{PP}$ & $A'$&$A'_{PP}$ & $A'-A'_{PP}$ \\
\hline
\multicolumn{9}{l} {}\\
\multicolumn{9}{l} {Sufficiently small $r_{12}$}\\
0.2&    0.46900&   0.44400&    0.197728&  0.196723&+0.001005&   0.118904& 0.077870&+0.041034\\
\hline

0.4&    0.43100&   0.42600&    0.386620&0.386538&+0.000082&   0.277525&      0.286919&-0.009394\\
\hline

0.6&    0.42620&   0.42280&    0.572037&   0.572068&-0.000031& 0.570774&      0.579790&-0.009016\\
\hline

0.8&    0.42800&   0.42280&    0.754729& 0.754721&+0.000008&  0.916385&      0.921649&-0.005264\\
\hline

1.0&    0.43110&   0.42460&    0.935242&   0.935213&+0.000029& 1.290323&     1.29312&-0.002797\\
\hline

2.0&    0.44550&   0.43680&    1.817203&   1.81720& +0.000003&3.347775&      3.34845&-0.000675\\
\hline
\multicolumn{9}{l} {}\\
\multicolumn{9}{l} {Sufficiently large $r_{12}$}\\
\hline
3.0&    0.45680&   0.44762&    2.680126&  2.68008&+0.000046& 5.536925&       5.53664&+0.000285\\
\hline

4.0&    0.46400&   0.45600&    3.533310&   3.53328&+0.000030&7.777719&       7.77753&+0.000189\\
\hline

5.0&    0.46870&   0.46235&    4.380874&  4.38088&-0.000006&10.045943&       10.0461&-0.000157\\
\hline

6.0&    0.47320&   0.46730&    5.224967& 5.22494&+0.000027& 12.331720&       12.3314&+0.000320\\
\hline

7.0&    0.47608&   0.47119&    6.066647&  6.06662&+0.000027&14.627793&       14.6276&+0.000193\\
\hline

8.0&    0.47820&   0.47430&    6.906648&  6.90665&-0.000002&16.931213&       16.9313&-0.000087\\
\hline

9.0&    0.48042&   0.47685&    7.745476&7.74546&+0.000016&  19.240602&       19.2405&+0.000102\\
\hline

10.0&   0.48202&   0.47898&    8.583369&  8.58335 &+0.000019&  21.553777&    21.5537&+0.000077\\
\hline

11.0&   0.48337&   0.48077&    9.420559&9.42055&+0.000009&  23.870068&       23.8700&+0.000068\\
\hline

12.0&   0.48460&   0.48230&   10.257201&  10.2572&+0.000001& 26.188828&      26.1888&+0.000028\\
\hline

13.0&   0.48553&   0.48356&   11.093399&  11.0934&-0.000001& 28.509499&      28.5095&-0.000001\\
\hline

14.0&   0.48661&   0.48471&   11.929252& 11.9292&+0.000052&  30.831947&      30.8318&+0.000147\\
\hline

15.0&   0.48720&   0.48570&   12.764781& 12.7648&-0.000019&  33.155366&      33.1554&-0.000034\\
\hline
\end{tabular}
\end{table}

\begin{table}[!ht]
\caption{ For the state $1s\sigma$ of the ion $peO^{8+}$ $(Z_1=1,
Z_2=8)$ the values of $C$ and $\tilde C$ have been obtained from the
requirement that the first- and third-order phase-integral results
coincide for $p$ as well as for $A'$. With the use of these values of
$C$ and $\tilde C$ the  values of $p$ and $A'$ have then been obtained
from the quantization conditions that are appropriate depending on
whether $r_{12}$ is sufficiently small or sufficiently large. The
numerically exact values obtained by Ponomarev and Puzynina [13] are
given in the columns called $p_{PP}$ and $A'_{PP}$.  }
\begin{tabular}{ccccccccc}
\hline
$r_{12}$ & $C$ & $\tilde C$ & $p$ & $p_{PP}$ & $p-p_{PP}$ & $A'$&$A'_{PP}$ & $A'-A'_{PP}$ \\
\hline
\multicolumn{9}{l} {}\\
\multicolumn{9}{l} {Sufficiently small $r_{12}$}\\
\hline
0.2&  0.5019300&  0.4809100&    0.857612&  0.855323&+0.002289& -0.137649&      -0.142128&+0.004479\\
\hline

0.4&  0.5039400&  0.4943700&    1.661131&  1.66144&+0.000087&  -0.366556&       -0.361348&-0.005208\\
\hline

0.6&  0.5038000&  0.4989000&    2.461700&   2.46187&-0.000170&  -0.576667&      -0.573808&-0.002859\\
\hline

0.8&  0.5033500&  0.5004300&    3.261920&  3.26197&-0.000050&  -0.782022&     -0.780407&-0.001615\\
\hline

1.0&  0.5028000&  0.5010000&    4.062060&  4.06206&0.00000&   -0.985331&     -0.984350&-0.000981\\
\hline

2.0&  0.5017000&  0.5012000&    8.062283&  8.06226&+0.000023&  -1.992407&     -1.99219&-0.000217\\
\hline
\multicolumn{9}{l} {}\\
\multicolumn{9}{l} {Sufficiently large $r_{12}$}\\
\hline
3.0&  0.5013000&  0.5010000&   12.062372&  12.0623&+0.000072&  -2.994818&     -2.99479&-0.000028\\ 
\hline

4.0&  0.5008300&  0.5007800&   16.062383&  16.0624&-0.000017&   -3.996148&    -3.99609&-0.000058\\
\hline
 
5.0&  0.5005000&  0.5007000&   20.062566&  20.0624&+0.000166&   -4.996555&     -4.99687&+0.000315\\
\hline
 
6.0&  0.5004300&  0.5005300&   24.062567&   24.0624&+0.000167&   -5.997131&   -5.99724&+0.000109\\
\hline

7.0&  0.5005000&  0.5005000&   28.062564&  28.0624&+0.000164&  -6.997498&     -6.99764&+0.000140\\
\hline

8.0&  0.5004000&  0.5004000&   32.062561&   32.0624&+0.000161&   -7.997818&   -7.99793&+0.000112\\
\hline

9.0&  0.5005000&  0.5004000&   36.062557&  36.0624&+0.000157&   -8.998017&    -8.99816&+0.000145\\
\hline

10.0&  0.5000000&  0.5004000&  40.062432&  40.0624&+0.000032&   -9.998458&    -9.99834&-0.000118\\
\hline

11.0&  0.5004000&  0.5003000&  44.062551& 44.0624&+0.000151&   -10.998401&   -10.9985&+0.000099\\
\hline

12.0&  0.5000000&  0.5003000&  48.062446&  48.0624&+0.000046&   -11.998716&  -11.9986&-0.000116\\
\hline

13.0&  0.5000000&  0.5003000&  52.062451& 52.0625&-0.000049&  -12.998804&    -12.9987&-0.000104\\
\hline

14.0&  0.5001000&  0.5003000&  56.062455& 56.0625&-0.000045&  -13.998861&    -13.9988&-0.000061\\
\hline

15.0&  0.4999900&  0.5002100&  60.062459& 60.0625&-0.000041&  -14.998981&      -14.9989&-0.000081\\
\hline
\end{tabular}
\end{table}

\begin{table}
\caption{
For the state $4d\sigma$ of the ion $peO^{8+}$ $(Z_1=1, Z_2=8)$ the values of $C$ and $\tilde C$ have been obtained
from the requirement that the first- and third-order phase-integral results
coincide for $p$ as well as for $A'$. With the use of these values of $C$ and $\tilde C$
the  values of $p$ and $A'$ have then been obtained from the 
quantization conditions that are appropriate depending
on whether $r_{12}$ is sufficiently small or sufficiently large. The numerically exact values
obtained by Ponomarev and Puzynina [13] are given in the columns called $p_{PP}$
and $A'_{PP}$.
}
\begin{tabular}{ccccccccc}
\hline
$r_{12}$ & $C$ & $\tilde C$ & $p$ & $p_{PP}$ & $p-p_{PP}$ & $A'$&$A'_{PP}$ & $A'-A'_{PP}$ \\
\hline
\multicolumn{9}{l} {}\\
\multicolumn{9}{l} {Sufficiently small $r_{12}$}\\
\hline
0.5&    0.6421800&    0.5329500&      0.567784&  0.569166& -0.001382& -6.45764&         -6.466968& +0.009328\\
\hline
1.0&    0.6301800&    0.5281010&      1.597590&   1.16192& -0.002161&  -7.813621&      -7.82220&   +0.008579\\   
\hline
2.0&   0.6223000&   0.5179000&     2.248680&  2.24564& +0.003040&-11.571664&      -11.5899&+0.018236\\
\hline
\multicolumn{9}{l} {}\\
\multicolumn{9}{l} {Sufficiently large $r_{12}$}\\
\hline

3.0&   0.5563400&   0.5069600&     3.260659&  3.26096&-0.000301& -15.145897&     -15.1384&-0.007497\\
\hline

4.0&   0.5434000&   0.5090000&     4.265089& 4.26518&-0.000091&   -18.414485&    -18.4123&-0.002185\\
\hline

5.0&   0.5357000 &  0.5114000&     5.265926&  5.26597&-0.000044&  -21.567077&    -21.5660&-0.001077\\
\hline

6.0&   0.5303000&   0.5127000&     6.265543&  6.26552&+0.000023&  -24.662029&    -24.6617&-0.000329\\
\hline

7.0&   0.5262000&   0.5131400&     7.264703&  7.26468&+0.000023&  -27.725910&    -27.7257&-0.000210\\
\hline

8.0&   0.5226000&   0.5131000&     8.263735&  8.26374&-0.000005&  -30.771316&    -30.7710&-0.000316\\
\hline

9.0&   0.5204500&   0.5127600&     9.262851&  9.26283&+0.000021&  -33.804409&    -33.8043&-0.000109\\
\hline

10.0&  0.5186100&   0.5123000&    10.261986&  10.2620&-0.000014&    -36.829957&  -36.8297&-0.000257\\
\hline

11.0&   0.5170000&  0.5118000&    11.261223&  11.2612&+0.000023&  -39.849843&    -39.8497&-0.000143\\
\hline

12.0&   0.5152000&  0.5113000&    12.260537&  12.2605&+0.000037&  -42.865768&    -42.8656&-0.000168\\
\hline

13.0&   0.5142000&  0.5107500&    13.259930&  13.2599&+0.000030&  -45.878744&    -45.8787&-0.000044\\
\hline

14.0&   0.5131800&  0.5102700&    14.259369&  14.2594& -0.000031&-48.889604&     -48.8896&-0.000004\\
\hline

15.0&   0.5122100&  0.5098200&    15.258859&  15.2589&-0.000051&  -51.898778&    -51.8987&-0.000078\\
\hline
\end{tabular}
\end{table}

\begin{table}
\caption{
For the state $1s\sigma$ of the ion $peHe^{2+}$ $(Z_1=1, Z_2=2)$ the values of $C$ and $\tilde C$ have been obtained
from the requirement that the first-order phase-integral results, obtained from quantization conditions that are appropriate depending
on whether $r_{12}$ is sufficiently small or sufficiently large, coincide for
$p$ as well as for $A'$ with the numerically exact results (accurate to all digits
quoted) calculated by Winter, Duncan and Lane [4], and obtained as private 
communication from Professor Winter (see p. 288 - 289 in [4]), and 
quoted  in this table as $p_{WDL}$ and $A'_{WDL}$.}
\begin{tabular}{ccccc}
\hline
   $r_{12}$& $p_{WDL}$  &   $ A'_{WDL}$&  $C$ & $\tilde C$  \\
\hline
\multicolumn{5}{l} {}\\
\multicolumn{5}{l} {Sufficiently small $r_{12}$}\\
\hline
0.2& 0.2909534228&  -0.0495531186& 0.51829036&  0.509271872\\
\hline
0.4& 0.5544040477& -0.1752443935& 0.532092811&  0.512039821\\
\hline
0.6& 0.7945061056& -0.3475381522& 0.552948218&  0.516402819\\
\hline
0.8& 1.018366017&  -0.5473748938& 0.562018492&  0.515629306\\
\hline
1.0& 1.231534107& -0.7624147481& 0.5529481616&  0.513927361\\
\hline
2.0& 2.24151& -1.86655& 0.5406495155& 0.5122852272\\ 
\hline
\multicolumn{5}{l} {}\\
\multicolumn{5}{l} {Sufficiently large $r_{12}$}\\
\hline
3.0& 3.241868168&-2.914992386 & 0.5148124860& 0.5120384660\\ 
\hline
4.0& 4.243211413& -3.937060587& 0.5060386160& 0.5106252670\\ 
\hline
5.0& 5.244326655&-4.949835242 & 0.5018822190& 0.5092751320\\ 
\hline
6.0& 6.245159553&-5.958257005 & 0.4994075600& 0.5081813590\\ 
\hline
7.0& 7.245789653&-6.964245417 & 0.4977688690& 0.5072953460\\ 
\hline
8.0&8.246278979 & -7.968726708& 0.4965834380& 0.5065545680\\ 
\hline
9.0& 9.246668544& -8.972207821& 0.4957006460& 0.5059509020\\  
\hline
10.0& 10.24698542& -9.974990618& 0.4950401840& 0.5054199100\\ 
\hline
11.0& 11.24724792& -10.97726635& 0.4943322500& 0.5050906930\\ 
\hline
12.0& 12.24746878& -11.97916218& 0.4940316000& 0.5045504220\\ 
\hline
13.0& 13.24765709& -12.98076598& 0.4934859900& 0.5044230599\\ 
\hline
14.0&14.24781949& -13.98214045& 0.4933158850& 0.5041397900\\ 
\hline
15.0& 15.24796097& -14.983331513& 0.4931492940& 0.5037628620\\ 
\hline
\end{tabular}
\end{table}

\begin{table}
\caption{
For the state $2p\sigma$ of the ion $peHe^{2+}$ $(Z_1=1, Z_2=2)$ the values of $C$ and $\tilde C$ have been obtained
from the requirement that the first-order phase-integral results, obtained from quantization conditions that are appropriate depending
on whether $r_{12}$ is sufficiently small or sufficiently large, coincide for
$p$ as well as for $A'$ with the numerically exact results (accurate to all digits
quoted) calculated by Winter, Duncan and Lane [4], and obtained as private 
communication from Professor Winter (see p. 288 - 289 in [4]), and 
quoted  in this table as $p_{WDL}$ and $A'_{WDL}$.}

\begin{tabular}{ccccc}
\hline
   $r_{12}$& $p_{WDL}$  &   $ A'_{WDL}$&  $C$ & $\tilde C$  \\
\hline
\multicolumn{5}{l} {}\\
\multicolumn{5}{l} {Sufficiently small $r_{12}$}\\
\hline
0.2 &0.1507994078& -2.013114367& 0.5419280238& 0.50829181\\
\hline
0.4& 0.3062680406& -2.053824124& 0.552930182&  0.50920298\\
\hline
0.6& 0.4698372597& -2.125935288& 0.572019821&  0.51028392\\
\hline
0.8& 0.6416038747& -2.234147173& 0.592038261&  0.52978182\\ 
\hline
1.0& 0.8180287700& -2.381560387& 0.619273017&  0.542910422\\
\hline
2.0& 1.640235157&  -3.846791567& 0.632937162&  0.569271625\\ 
\hline
\multicolumn{5}{l} {}\\
\multicolumn{5}{l} {Sufficiently large $r_{12}$}\\
\hline
3.0& 2.303194434&  -5.444185235& 0.65293027&   0.551927319\\
\hline
4.0& 2.872046343&  -7.448941809& 0.6829729310& 0.549321938\\
\hline
5.0& 3.395848335& -9.526950457& 0.7689678921& 0.5473760702\\ 
\hline
6.0& 3.901954918& -11.61569923& 0.6760630853& 0.5461458940\\ 
\hline
7.0& 4.404864367& -13.68696569& 0.6153175050& 0.5440702840\\ 
\hline
8.0& 4.909225930& -15.73665989& 0.5787297600& 0.5416277770\\ 
\hline
9.0& 5.414760984& -17.77106869& 0.5582392230& 0.5391808520\\ 
\hline
10.0& 5.920460066& -19.79629610& 0.5461971300& 0.5368472710\\ 
\hline
11.0& 6.425795942& -21.81592389& 0.5382876680& 0.5347039250\\ 
\hline
12.0& 6.930614764& -23.83184680& 0.5325850510& 0.5327246020\\ 
\hline
13.0& 7.434919206& -25.84512132& 0.5281541630& 0.5308665610\\ 
\hline
14.0& 7.938760040& -27.85639810& 0.5245807700& 0.5291925630\\ 
\hline
15.0& 8.442196146& -29.86611405& 0.5216764050& 0.5276965330\\ 
\hline
\end{tabular}

\end{table}

\begin{table}
\caption{
For the state $1s\sigma$ of the ion $peB^{5+}$ $(Z_1=1, Z_2=5)$ the values of $C$ and $\tilde C$ have been obtained
from the requirement that the first-order phase-integral results, obtained from quantization conditions that are appropriate depending
on whether $r_{12}$ is sufficiently small or sufficiently large, coincide for
$p$ as well as for $A'$ with the numerically exact results obtained by 
Ponomarev and Puzynina [13] and 
quoted  in this table as $p_{PP}$ and $A'_{PP}$.}
\begin{tabular}{ccccc}
\hline
   $r_{12}$& $p_{PP}$  &   $ A'_{PP}$&  $C$ & $\tilde C$  \\
\hline
\multicolumn{5}{l} {}\\
\multicolumn{5}{l} {Sufficiently small $r_{12}$}\\
\hline
     0.2&      0.57180&        -0.106339&  0.525368&    0.501193\\
\hline
     0.4&      1.09319&   -0.312917&    0.523294&    0.500928\\   
\hline
    0.6&       1.59754&   -0.533666&    0.521226&    0.500636\\
\hline
     0.8&      2.09827&   -0.749331&    0.516453&    0.503010\\
\hline
     1.0&      2.59847&   -0.959583&    0.512521&    0.503770\\
\hline
     2.0&      5.09906&   -1.97997&    0.505470&    0.503400\\
\hline
\multicolumn{5}{l} {}\\
\multicolumn{5}{l} {Sufficiently large $r_{12}$}\\
\hline
     3.0&      7.59936&   -2.98666&    0.503530&    0.502600\\
\hline
     4.0&      10.0995&   -3.99000&    0.502580&    0.502110\\
\hline
     5.0&      12.5996&   -4.99200&    0.502040&    0.501740\\
\hline
    6.0&       15.0997&   -5.99333&    0.501760&    0.501410\\
\hline
     7.0&      17.5997&   -6.99428&    0.501420&    0.501320\\
\hline
     8.0&      20.0998&   -7.99500&    0.501350&    0.501030\\
\hline
     9.0&      22.5998&   -8.99556&    0.501140&    0.501000\\
\hline
    10.0&      25.0998&   -9.99563&    0.501400&    0.501300\\
\hline
    11.0&      27.5998&   -10.9960&    0.501220&    0.501230\\
\hline
    12.0&      30.0998&   -11.9964&    0.501010&    0.501100\\
\hline
    13.0&      32.5998&   -12.9966&    0.501000&    0.501110\\
\hline
    14.0&      35.0998&   -13.9969&    0.500810&    0.501000\\
\hline
    15.0&      37.5999&   -14.9971&    0.501000&    0.500800\\
\hline
\end{tabular}
\end{table}

\begin{table}
\caption{
For the state $3s\sigma$ of the ion $peB^{5+}$ $(Z_1=1, Z_2=5)$ the values of $C$ and $\tilde C$ have been obtained
from the requirement that the first-order phase-integral results, obtained from quantization conditions that are appropriate depending
on whether $r_{12}$ is sufficiently small or sufficiently large, coincide for
$p$ as well as for $A'$ with the numerically exact results obtained by 
Ponomarev and Puzynina [13] and
quoted  in this table as $p_{PP}$ and $A'_{PP}$.}
\begin{tabular}{ccccc}
\hline
   $r_{12}$& $p_{PP}$  &   $ A'_{PP}$&  $C$ & $\tilde C$  \\
\hline
\multicolumn{5}{l} {}\\
\multicolumn{5}{l} {Sufficiently small $r_{12}$}\\
\hline
0.2&      0.196723&    0.077870&    0.428610&    0.440763\\
\hline
0.4&      0.386538&    0.286919&    0.440847&   0.437841\\
\hline
0.6&      0.572068&    0.579790&    0.436194&    0.431006\\
\hline
0.8&      0.754721&    0.921649&    0.434223&    0.428242\\
\hline
1.0&      0.935213&    1.29132&    0.433012&    0.426215\\
\hline
2.0&      1.81720&    3.34845&    0.446970&   0.437520\\
\hline
\multicolumn{5}{l} {}\\
\multicolumn{5}{l} {Sufficiently large $r_{12}$}\\
\hline
3.0&      2.68008&    5.53664&    0.457260&    0.448050\\
\hline
4.0&      3.53328&    7.77753&    0.464410&    0.456250\\
\hline
5.0&      4.38088&    10.0461&   0.469530&    0.462430\\
\hline
6.0&      5.22494&    12.3314&    0.473400&    0.467340\\
\hline
7.0&      6.06662&    14.6276&   0.476400&    0.471350\\
\hline
8.0&      6.90665&    16.9313&    0.478790&    0.474360\\
\hline
9.0&      7.74546&    19.2405&    0.480780&    0.476950\\
\hline
10.0&      8.58335&    21.5537&   0.482370&    0.479130\\
\hline
11.0&      9.42055&    23.8700&    0.483710&    0.480810\\
\hline
12.0&      10.2572&    26.1888&    0.484920&    0.482270\\
\hline
13.0&      11.0934&    28.5095&    0.485910&    0.483550\\
\hline
14.0&      11.9292&    30.8318&    0.486720&    0.485160\\
\hline
15.0&      12.7648&    33.1554&    0.487600&    0.485520\\
\hline
\end{tabular}

\end{table}

\begin{table}

\caption{For the state $1s\sigma$ of the ion $peO^{8+}$ $(Z_1=1,
Z_2=8)$ the values of $C$ and $\tilde C$ have been obtained from the
requirement that the first-order phase-integral results, obtained from
quantization conditions that are appropriate depending on whether
$r_{12}$ is sufficiently small or sufficiently large, coincide for $p$
as well as for $A'$ with the numerically exact results obtained by
Ponomarev and Puzynina [13] and quoted  in this table as $p_{PP}$ and
$A'_{PP}$.}

\begin{tabular}{ccccc}
\hline
   $r_{12}$& $p_{PP}$  &   $ A'_{PP}$&  $C$ & $\tilde C$  \\
\hline
\multicolumn{5}{l} {}\\
\multicolumn{5}{l} {Sufficiently small $r_{12}$}\\
\hline
      0.2&      0.855323&   -0.142128&    0.494838&    0.483907\\
\hline
      0.4&      1.66144&   -0.361348&    0.509680&    0.498740\\
\hline
      0.6&      2.46187&   -0.573808&    0.506980&    0.501340\\
\hline
      0.8&      3.26197&   -0.780407&   0.505080&    0.501930\\
\hline
      1.0&      4.06206&   -0.984350&    0.504010&    0.501980\\
\hline
      2.0&      8.06226&   -1.99219&    0.501950&    0.501460\\  
\hline
\multicolumn{5}{l} {}\\
\multicolumn{5}{l} {Sufficiently large $r_{12}$}\\
\hline
      3.0&      12.0623&   -2.99479&    0.501220&    0.501160\\
\hline
      4.0&      16.0624&   -3.99609&    0.501010&    0.500800\\
\hline
      5.0&      20.0624&   -4.99687&    0.500800&    0.500700\\
\hline
      6.0&      24.0624&   -5.99724&    0.500800&    0.500800\\
\hline
      7.0&      28.0624&   -6.99764&    0.500600&    0.500700\\
\hline
      8.0&      32.0624&   -7.99793&    0.500500&   0.500620\\
\hline
      9.0&      36.0624&   -8.99816&    0.500410&    0.500600\\
\hline
     10.0&      40.0624&   -9.99834&    0.500400&    0.500600\\
\hline
     11.0&      44.0624&   -10.9985&    0.500300&    0.500500\\
\hline
     12.0&      48.0624&   -11.9986&    0.500300&    0.500500\\
\hline
    13.0&      52.0625&   -12.9987&    0.500500&    0.500300\\
\hline
     14.0&      56.0625&   -13.9988&    0.500400&    0.500300\\
\hline
     15.0&      60.0625&   -14.9989&    0.500400&    0.500200\\
\hline
\end{tabular}
\end{table}

\begin{table}
\caption{
For the state $4d\sigma$ of the ion $peO^{8+}$ $(Z_1=1, Z_2=8)$ the values of $C$ and $\tilde C$ have been obtained
from the requirement that the first-order phase-integral results, obtained from quantization conditions that are appropriate depending
on whether $r_{12}$ is sufficiently small or sufficiently large, coincide for
$p$ as well as for $A'$ with the numerically exact results obtained by 
Ponomarev and Puzynina [13] and
quoted  in this table as $p_{PP}$ and $A'_{PP}$.}

\begin{tabular}{ccccc}
\hline
   $r_{12}$& $p_{PP}$  &   $ A'_{PP}$&  $C$ & $\tilde C$  \\
\hline
\multicolumn{5}{l} {}\\
\multicolumn{5}{l} {Sufficiently small $r_{12}$}\\
\hline
     0.5&      0.569166&   -6.466968&    0.639180&     0.552970\\  
\hline
     1.0&      1.16192&    -7.82220&     0.628280&   0.547102\\  
\hline
     2.0&     2.24564&   -11.5899&    0.594960&     0.535600\\ 
\hline
\multicolumn{5}{l} {}\\
\multicolumn{5}{l} {Sufficiently large $r_{12}$}\\
\hline
     3.0&     3.26096&   -15.1384&    0.565392&     0.511310\\ 
\hline
     4.0&      4.26518&   -18.4123&    0.546141&     0.510320\\ 
\hline
     5.0&      5.26597&   -21.5660&    0.537100&     0.512080\\ 
\hline
     6.0&      6.26552&   -24.6617&    0.531130&     0.513220\\ 
\hline
     7.0&      7.26468&   -27.7257&    0.526900&     0.513530\\ 
\hline
     8.0&      8.26374&   -30.7710&    0.523520&     0.513370\\ 
\hline
     9.0&      9.26283&   -33.8043&    0.520970&     0.513030\\ 
\hline
    10.0&      10.2620&   -36.8297&    0.519020&     0.512460\\ 
\hline
    11.0&     11.2612&   -39.8497&    0.516950&     0.512100\\ 
\hline
    12.0&      12.2605&   -42.8656&    0.515510&     0.511710\\ 
\hline
    13.0&      13.2599&   -45.8787&    0.514440&     0.511010\\ 
\hline
    14.0&      14.2594&  -48.8896&    0.513870&     0.510060\\ 
\hline
    15.0&      15.2589&   -51.8987&    0.513060&     0.509620\\ 
\hline
\end{tabular}
\end{table}

\begin{figure}[!ht]
\centerline{\epsfig{figure=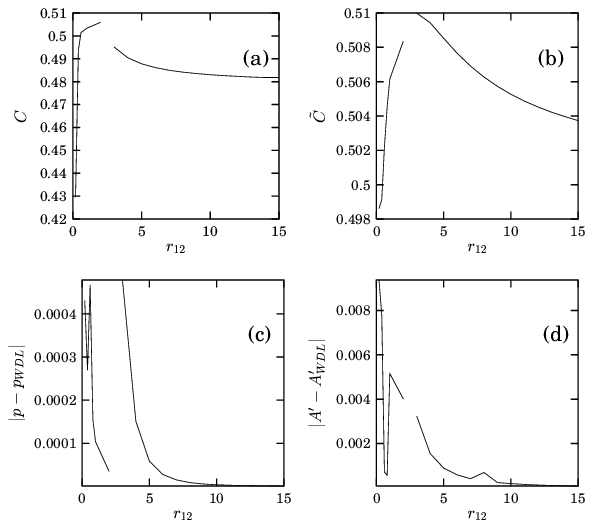, width=\linewidth}}
\caption
{ Plots for the $1s\sigma$ state of the ion $peHe^{2+}$ $(Z_1=1, Z_2=2)$ of (a) $C$ versus $r_{12}$, 
(b) $\tilde C$ versus $r_{12}$, (c) $|p-p_{WDL}|$ 
versus $r_{12}$ and (d) $|A'-A'_{WDL}|$ versus $r_{12}$, 
when $C$ and $\tilde C$ are determined 
as functions of $r_{12}$ from the
requirement that the first-order and the third-order phase-integral 
results for $p$ as well as for $A'$ coincide. Here $p$ and $A'$ are the phase-integral values obtained in
Table I, while $p_{WDL}$ and $A'_{WDL}$ are the corresponding numerically
exact values (accurate to all digits quoted) calculated by Winter, Duncan and Lane [4] (see p. 288-289 in [4]),
and quoted in Table I. There is a break in each curve between the regions where
the quantization conditions for sufficiently small and sufficiently large values
of $r_{12}$ have been used. }
\label{Fig.3(a)}
\end{figure}

\begin{figure}[!ht]
\centerline{\epsfig{figure=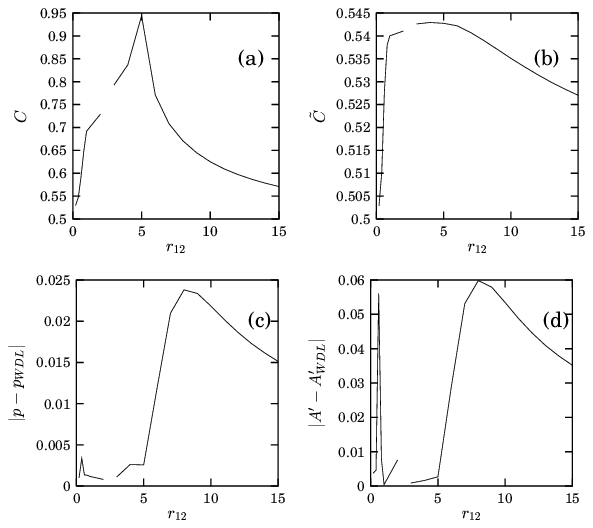, width=\linewidth}}
\caption
{ Plots for the $2p\sigma$ state of the ion $peHe^{2+}$ $(Z_1=1, Z_2=2)$ of (a) $C$ versus $r_{12}$, 
(b) $\tilde C$ versus $r_{12}$, (c) $|p-p_{WDL}|$ 
versus $r_{12}$ and (d) $|A'-A'_{WDL}|$ versus $r_{12}$, 
when $C$ and $\tilde C$ are determined 
as functions of $r_{12}$ from the
requirement that the first-order and the third-order phase-integral 
results for $p$ as well as for $A'$ coincide. Here $p$ and $A'$ are the phase-integral values obtained in
Table II, while $p_{WDL}$ and $A'_{WDL}$ are the corresponding numerically
exact values (accurate to all digits quoted) calculated by Winter, Duncan and Lane [4] (see p. 288-289 in [4]),
and quoted in Table II. There is a break in each curve between the regions where
the quantization conditions for sufficiently small and sufficiently large values
of $r_{12}$ have been used.}
\label{Fig.3(b)}
\end{figure}

\begin{figure}[!ht]
\centerline{\epsfig{figure=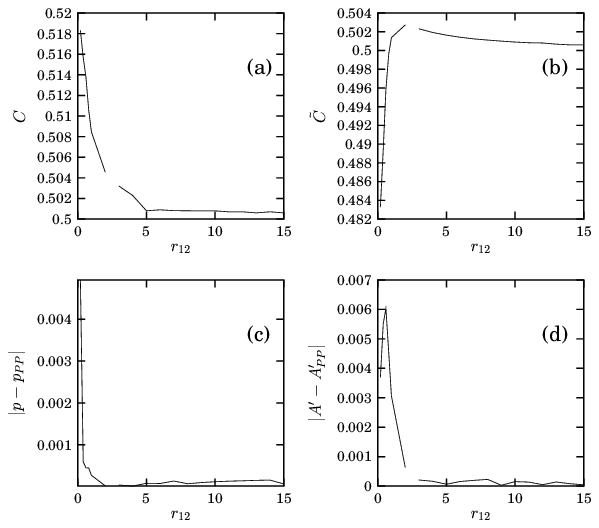, width=\linewidth}}
\caption
{ Plots for the $1s\sigma$ state of the ion $peB^{5+}$ $(Z_1=1, Z_2=5)$ of (a) $C$ versus $r_{12}$, 
(b) $\tilde C$ versus $r_{12}$, (c) $|p-p_{PP}|$ 
versus $r_{12}$ and (d) $|A'-A'_{PP}|$ versus $r_{12}$, 
when $C$ and $\tilde C$ are determined 
as functions of $r_{12}$ from the
requirement that the first-order and the third-order phase-integral 
results for $p$ as well as for $A'$ coincide. Here $p$ and $A'$ are the phase-integral values obtained in
Table III, while $p_{PP}$ and $A'_{PP}$ are the corresponding numerically
exact values obtained by Ponomarev and Puzynina [13] and quoted Table III. 
There is a break in each curve between the regions where
the quantization conditions for sufficiently small and sufficiently large values
of $r_{12}$ have been used.}
\label{Fig.3(c)}
\end{figure}

\begin{figure}[!ht]
\centerline{\epsfig{figure=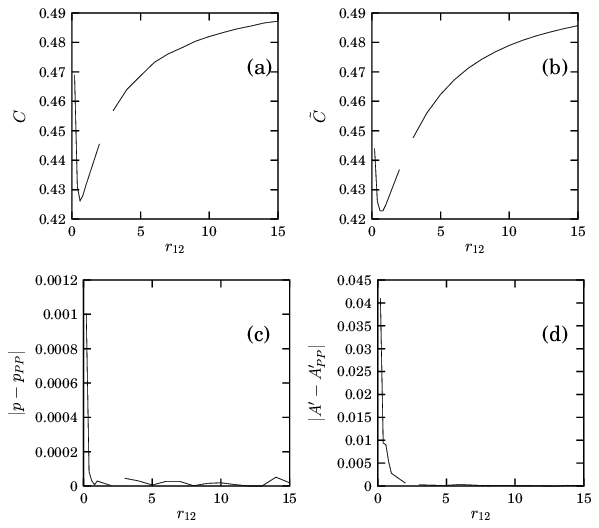, width=\linewidth}}
\caption
{ Plots for the $3s\sigma$ state of the ion $peB^{5+}$ $(Z_1=1, Z_2=5)$ of (a) $C$ versus $r_{12}$, 
(b) $\tilde C$ versus $r_{12}$, (c) $|p-p_{PP}|$ 
versus $r_{12}$ and (d) $|A'-A'_{PP}|$ versus $r_{12}$, 
when $C$ and $\tilde C$ are determined 
as functions of $r_{12}$ from the
requirement that the first-order and the third-order phase-integral 
results for $p$ as well as for $A'$ coincide. Here $p$ and $A'$ are the phase-integral values obtained in
Table IV, while $p_{PP}$ and $A'_{PP}$ are the corresponding numerically
exact values obtained by Ponomarev and Puzynina [13] and quoted in Table IV.
There is a break in each curve between the regions where
the quantization conditions for sufficiently small and sufficiently large values
of $r_{12}$ have been used.}
\label{Fig.3(d)}
\end{figure}

\begin{figure}[!ht]
\centerline{\epsfig{figure=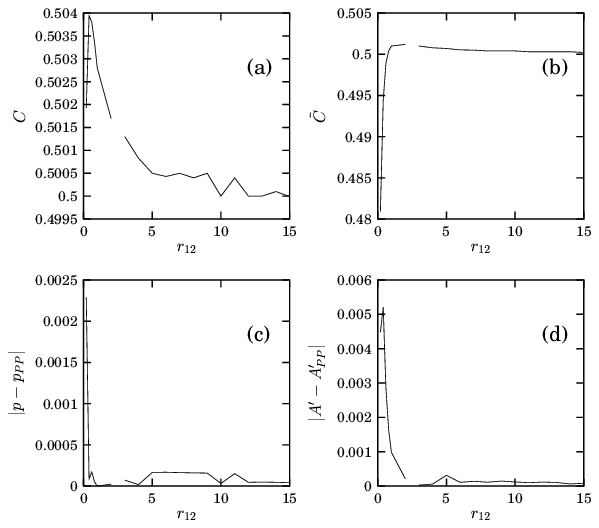, width=\linewidth}}
\caption
{ Plots for the $1s\sigma$ state of the ion $peO^{8+}$ $(Z_1=1, Z_2=8)$ of (a) $C$ versus $r_{12}$, 
(b) $\tilde C$ versus $r_{12}$, (c) $|p-p_{PP}|$ 
versus $r_{12}$ and (d) $|A'-A'_{PP}|$ versus $r_{12}$, 
when $C$ and $\tilde C$ are determined 
as functions of $r_{12}$ from the
requirement that the first-order and the third-order phase-integral 
results for $p$ as well as for $A'$ coincide. Here $p$ and $A'$ are the phase-integral values obtained in
Table V, while $p_{PP}$ and $A'_{PP}$ are the corresponding numerically
exact values obtained by Ponomarev and Puzynina [13] and quoted in Table V.
There is a break in each curve between the regions where
the quantization conditions for sufficiently small and sufficiently large values
of $r_{12}$ have been used.}
\label{Fig.3(e)}
\end{figure}

\begin{figure}[!ht]
\centerline{\epsfig{figure=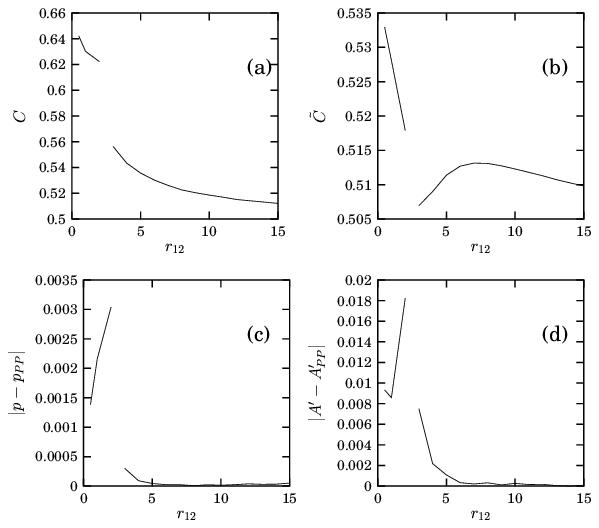, width=\linewidth}}
\caption
{ Plots for the $4d\sigma$ state of the ion $peO^{8+}$ $(Z_1=1, Z_2=8)$ of (a) $C$ versus $r_{12}$, 
(b) $\tilde C$ versus $r_{12}$, (c) $|p-p_{PP}|$ 
versus $r_{12}$ and (d) $|A'-A'_{PP}|$ versus $r_{12}$, 
when $C$ and $\tilde C$ are determined 
as functions of $r_{12}$ from the
requirement that the first-order and the third-order phase-integral 
results for $p$ as well as for $A'$ coincide. Here $p$ and $A'$ are the phase-integral values obtained in
Table VI, while $p_{PP}$ and $A'_{PP}$ are the corresponding numerically
exact values obtained by Ponomarev and Puzynina [13] and quoted in Table VI.
There is a break in each curve between the regions where
the quantization conditions for sufficiently small and sufficiently large values
of $r_{12}$ have been used.}
\label{Fig.3(f)}
\end{figure}

\begin{figure}[!ht]
\centerline{\epsfig{figure=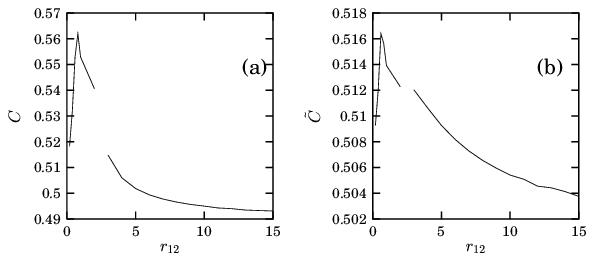, width=\linewidth}}
\caption{ Plots for the $1s\sigma$ state of the ion $peHe^{2+}$ $(Z_1=1, Z_2=2)$ 
of (a) $C$ versus $r_{12}$ and (b) $\tilde C$  versus $r_{12}$, 
 when $C$ and $\tilde C$ are determined 
as functions of $r_{12}$ from the
requirement that the first-order phase-integral results 
and the numerically
exact results obtained by Winter {\it et al.} [4] coincide. 
There is a break in each curve between the regions where
the quantization conditions for sufficiently small and sufficiently large values
of $r_{12}$ have been used.}   
\label{Fig.3(g)}
\end{figure}

\begin{figure}[!ht]
\centerline{\epsfig{figure=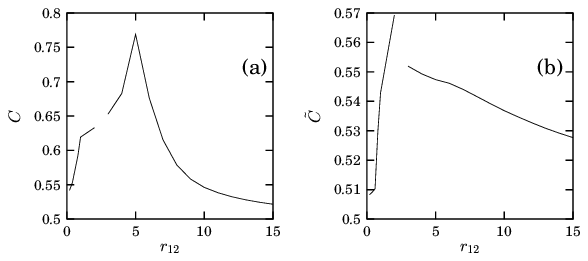, width=\linewidth}}
\caption{ Plots for the $2p\sigma$ state of the ion $peHe^{2+}$ $(Z_1=1, Z_2=2)$ of 
(a) $C$ versus $r_{12}$ and (b) $\tilde C$  versus $r_{12}$, 
when $C$ and $\tilde C$ are determined 
as functions of $r_{12}$ from the
requirement that the first-order phase-integral results and the numerically
exact results obtained by Winter {\it et al.} [4] coincide.
There is a break in each curve between the regions where
the quantization conditions for sufficiently small and sufficiently large values
of $r_{12}$ have been used.}
\label{Fig.3(h)} 
\end{figure}

\begin{figure}[!ht]
\centerline{\epsfig{figure=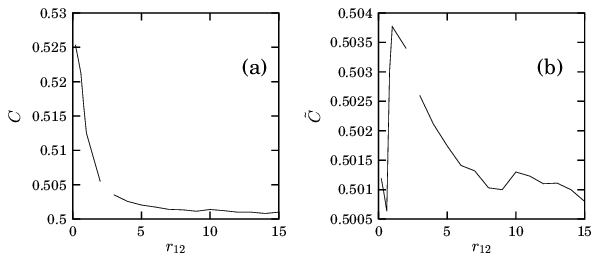, width=\linewidth}}
\caption{ Plots for the $1s\sigma$ state of the ion 
$peB^{5+}$ $(Z_1=1, Z_2=5)$ of (a) $C$ versus $r_{12}$ and (b) $\tilde C$  
versus $r_{12}$, 
 when $C$ and $\tilde C$ are determined 
as functions of $r_{12}$ from the
requirement that the first-order phase-integral results and the numerically
exact results obtained by Ponomarev and Puzynina [13] coincide.
There is a break in each curve between the regions where
the quantization conditions for sufficiently small and sufficiently large values
of $r_{12}$ have been used.}
\label{Fig.3(i)}
\end{figure}

\begin{figure}[!ht]
\centerline{\epsfig{figure=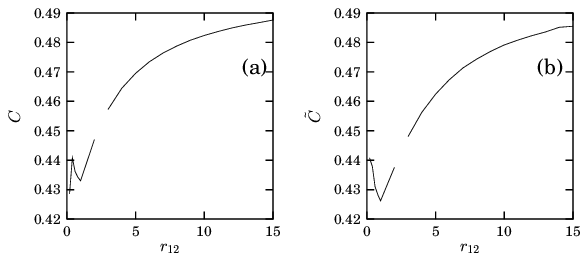, width=\linewidth}}
\caption{ Plots for the $3s\sigma$ state of the ion 
$peB^{5+}$ $(Z_1=1, Z_2=5)$ of (a) $C$ versus $r_{12}$ and (b) $\tilde C$ 
 versus $r_{12}$, 
when $C$ and $\tilde C$ are determined 
as functions of $r_{12}$ from the
requirement that the first-order phase-integral results and the numerically
exact results obtained by Ponomarev and Puzynina [13] coincide.
There is a break in each curve between the regions where
the quantization conditions for sufficiently small and sufficiently large values
of $r_{12}$ have been used.}
\label{Fig.3(j)}
\end{figure}

\begin{figure}[!ht]
\centerline{\epsfig{figure=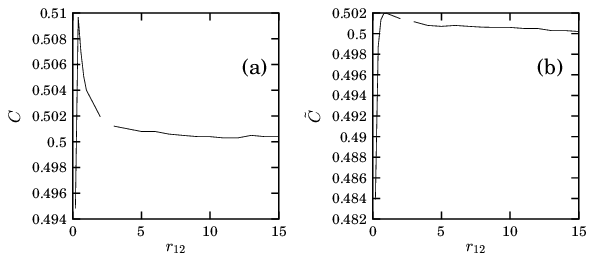, width=\linewidth}}
\caption{ Plots for the $1s\sigma$ state of the ion 
$peO^{8+}$ $(Z_1=1, Z_2=8)$ of (a) $C$ versus $r_{12}$ and (b) $\tilde C$ 
 versus $r_{12}$, 
when $C$ and $\tilde C$ are determined 
as functions of $r_{12}$ from the
requirement that the first-order phase-integral results and the numerically
exact results obtained by Ponomarev and Puzynina [13] coincide.
There is a break in each curve between the regions where
the quantization conditions for sufficiently small and sufficiently large values
of $r_{12}$ have been used.}
\label{Fig.3(k)}
\end{figure}

\begin{figure}[!ht]
\centerline{\epsfig{figure=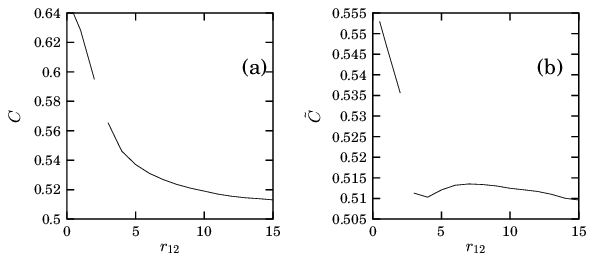, width=\linewidth}}
\caption{ Plots for the $4d\sigma$ state of the ion 
$peO^{8+}$ $(Z_1=1, Z_2=8)$ of (a) $C$ versus $r_{12}$ and (b) $\tilde C$  
versus $r_{12}$, 
 when $C$ and $\tilde C$ are determined 
as functions of $r_{12}$ from the
requirement that the first-order phase-integral results and the numerically
exact results obtained by Ponomarev and Puzynina [13] coincide.
There is a break in each curve between the regions where
the quantization conditions for sufficiently small and sufficiently large values
of $r_{12}$ have been used.}
\label{Fig.3(l)}
\end{figure}
\end{document}